\documentclass[preprint,12pt,authoryear]{elsarticle}
\usepackage{lineno}
\usepackage[fleqn]{amsmath}
\usepackage{etoolbox} 
\newcommand*\linenomathpatch[1]{%
  \cspreto{#1}{\linenomath}%
  \cspreto{#1*}{\linenomath}%
  \csappto{end#1}{\endlinenomath}%
  \csappto{end#1*}{\endlinenomath}%
}

\linenomathpatch{equation}
\linenomathpatch{gather}
\linenomathpatch{multline}
\linenomathpatch{align}
\linenomathpatch{alignat}
\linenomathpatch{flalign}

\usepackage{amssymb}
\usepackage{epstopdf}
\usepackage{ulem}
\usepackage{color}
\usepackage{hyperref}

\DeclareMathOperator{\sech}{sech}

\journal{Dynamics of Atmospheres and Oceans}

\begin{document}

\begin{frontmatter}

\title{Effects of equatorially-confined shear flow on MRG and Rossby waves}

\author[ICTS]{Mukesh Singh Raghav}
\affiliation[ICTS]{organization={International Centre for Theoretical Sciences, Tata Institute of Fundamental Research},
            city={Bengaluru},
            postcode={$560089$}, 
            country={India}}
\affiliation[iiser]{organization={Department of Data Science, Indian Institute of Science Education and Research},
            city={Pune},
            postcode={$411008$}, 
            country={India}}
 
\cortext[aa_cor]{Corresponding author}            
\author[ICTS]{Sharath Jose}
\author[ICTS,iiser]{Amit Apte\corref{aa_cor}}
\ead{apte@iiserpune.ac.in}
\author[ICTS]{Rama Govindarajan}

\begin{abstract}
Linear modal stability analysis of a mean zonal shear flow is carried out in the framework of rotating shallow water equations (RSWE), both under the $\beta$-plane approximation and in the full spherical coordinate system. Two base flows -- equatorial easterly (EE) and westerly (EW) -- with Gaussian profiles highly confined to small latitudes are analyzed. At low Froude number, mixed Rossby-gravity (MRG) and Rossby waves are found to be particularly affected by shear, with prominent changes at higher wavenumbers. These waves become practically non-dispersive at large wavenumbers in EE. The perturbations are found to be more confined equatorially in EE than in EW with the degree of confinement being more pronounced in the $\beta$-plane system compared to the full spherical system. At high Froude number, the phase speeds are significantly larger in the $\beta$-plane system for all families of waves. Under the $\beta$-plane approximation, exponentially unstable modes can be excited, having negative (positive) phase speed in EE (EW). Strikingly, this flow is always neutrally stable with the full spherical system. This speaks for the importance of studying the whole spherical system even for equatorially confined shear. 
\end{abstract}
\begin{graphicalabstract}
\begin{figure}
  \includegraphics[width=1\textwidth]{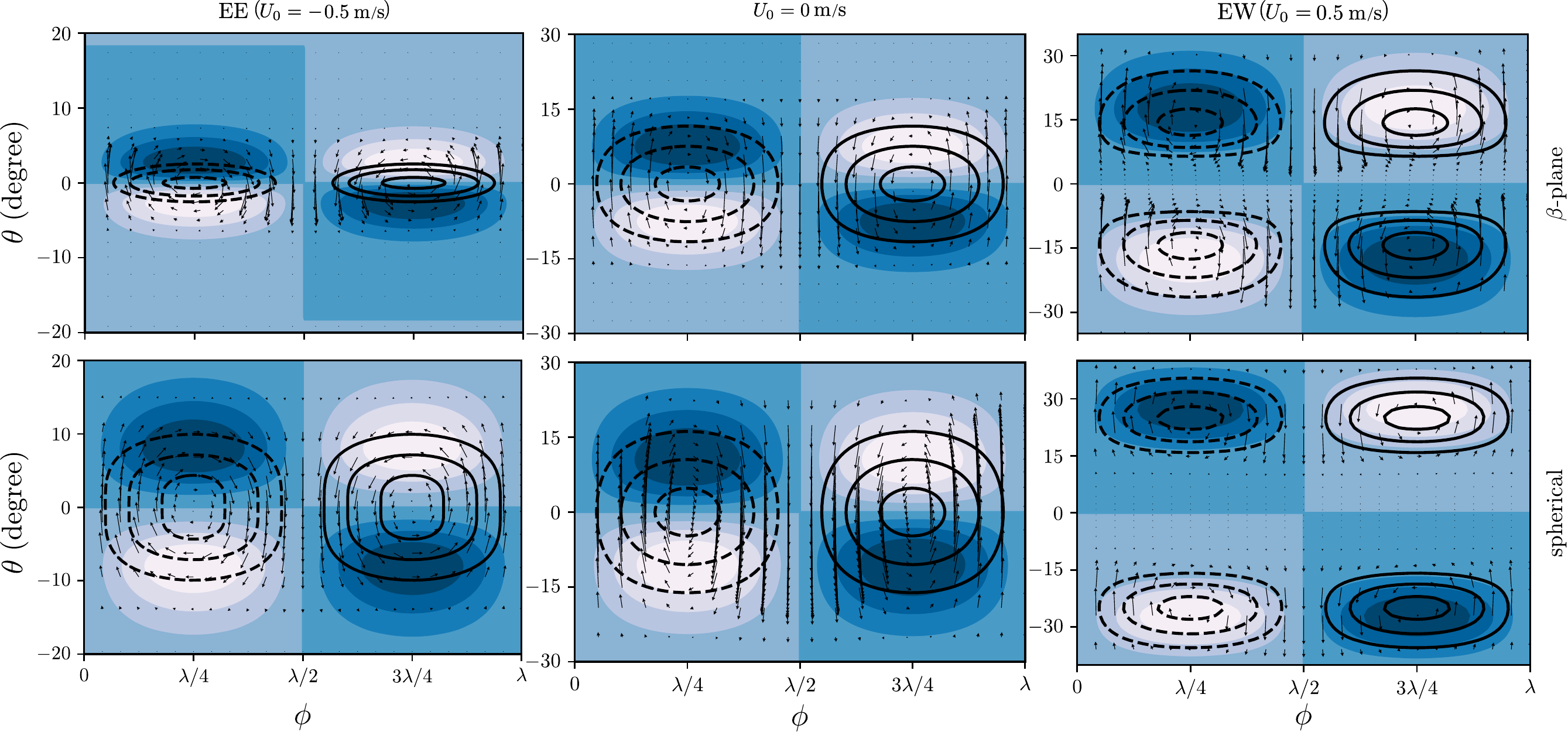}
\end{figure}
The structure of equatorial waves is greatly affected by the presence of shear flow. When compared with the zero mean flow case, these waves are more (less) equatorially trapped for an easterly (westerly) mean flow. In the $\beta$-plane setting, the meridional extent of the waves is underestimated even for highly equatorially confined mean flows. Our study thus highlights the discrepancies in the wave characteristics even under favourable conditions for comparison between the $\beta$-plane and the spherical systems.

\end{graphicalabstract}

\begin{highlights}
\item Stability analysis of an equatorially confined shear flow is conducted using beta plane and full spherical equations.
\item The structures of equatorial waves are shown to differ from the classical waves without shear.
\item It is shown that even for shear which is highly equatorially confined, the resulting atmospheric waves are not.
\item The full spherical stability analysis provides qualitatively different waves from the waves in beta plane theory.
\end{highlights}


\end{frontmatter}

\setcounter{figure}{0}
\section{Introduction}
The theory of equatorial waves is crucial for an understanding of equatorial dynamics in geophysical systems~\citep{kiladis2009}. In a seminal work, \cite{Matsuno} derived analytical solutions to the rotating shallow water equations (RSWE) on an equatorial $\beta$-plane linearised about a steady state with zero flow. Matsuno's analysis identified different families of waves -- eastward inertio-gravity (EIG), westward inertio-gravity (WIG), mixed Rossby gravity (MRG), Rossby and Kelvin waves. Subsequent observations established the presence of these waves in the atmosphere and the ocean~\citep{Yanai1966,Lindzen1968QBO,Wallace1968,Holton1968kelvin,takayabu1994a}.

These equatorial waves are known to have notable effects on the dynamics of the atmosphere and the ocean~\citep{vallis_2017,boyd2018dynamics}. For instance, the modification of the zonally symmetric flow in the stratosphere is intimately linked to the MRG wave~\citep{wallace1973}. Further, Kelvin waves are known to generate an upward flux of westerly momentum, accounting for westerly acceleration associated with the quasi-biennial oscillation~\citep{Wallace1968}. The agreement of the prominent peaks in the global space-time spectrum of the tropical cloudiness~\citep{WK99} and the rainfall~\citep{Cho04} with theoretically calculated dispersion curves has proved to be a remarkable success for the theory. 

The Matsuno theory is based on zero mean flow, but meridionally averaged climatological zonal flow shows significant variation with latitude and presence of longitudinal shear. Thus a natural question is to understand how the equatorial waves are modified in the presence of such shear flows - this is one of main aims of the present study. We first review some of the main results of previous studies in this direction.

Beyond the primary effects of Doppler shifting~\citep{yang2003}, mean flow is shown to have non-trivial effects on the properties of equatorial waves and vice versa~\citep{dias2014,buhler2014}. Critical layers, where the phase speed equals the mean flow velocity and disturbance energy is absorbed, can exist~\citep{bennett1971}. The longitudinal variation of the mean flow can result in regions where wave energy can be accumulated or depleted; here, wave trains extending beyond the equatorial regions can allow for energy exchange with higher latitudes~\citep{websterchang1988}. MRG and Rossby modes show a higher degree of trapping in the presence of equatorial easterly than in a westerly mean flow ~\citep{WZ89}.

While the $\beta$-plane approximation often yields good insights, it must be appreciated that the dynamics on the sphere can assume greater complexity~\citep{hough1898}. In a landmark work, \cite{LH68} described in great detail the numerically computed eigenfunctions on the sphere for the complete range of $\epsilon$ (a parameter characterising rotation that we define in section \ref{sec:gov_eq}), and further discussed the asymptotic forms of these eigenfunctions in various limits of $\epsilon$ that revealed interesting properties of gravity, planetary and Kelvin waves.
Recent work~\citep{paldor2018MRG} showed the solutions in the weak rotation regime to be suitable for most wavenumbers while the solutions in the strong rotation regime seem to work well for only small wavenumbers. The Kelvin mode, which was identified as a distinct mode in the $\beta$-plane configuration, was found to be the lowest order EIG mode in the spherical system~\citep{paldor2017Kelvin}. In the presence of strong polar and equatorial jets, \cite{paldor2021} found the shallow layer depth to be a crucial factor for the growth rates of modal instabilities.

The following two comprehensive studies of stability of RSWE with a longitudinal shear flow are most relevant to the present paper: \cite{WZ89} using the $\beta$-plane approximation, and \cite{paldor2021} using the full spherical system. \cite{WZ89} studied the case of a mid-latitude jet in addition to equatorial easterly (EE) or westerly (EW) jets. These profiles have features of a realistic flow, but they extend beyond the equatorial region. So in principle, such flows should be studied in a complete spherical system rather than under the $\beta$-plane approximation. Also, this early study was restricted to small wavenumbers. As will be apparent in our study, the effects of shear flow are significant for large wavenumbers, which is one of the main contributions of our work.

\cite{paldor2021} studied the stability of equatorial jets in the spherical system. Their motivation was to compare the instabilities obtained from approximate theories -- non-divergent (ND) and quasi-geostrophic (QG) -- to those from the full shallow water equations. They showed that the full equations consistently predict smaller disturbance growth rates. We note that their flow speeds were very large, whereas we study weaker shears which occur more commonly in the atmosphere. We note further that their work was primarily interested in the growth rates of unstable modes, and did not examine the spatial structure of the perturbations, which is another major contribution of the present paper.

With this background in mind, we now review the salient features and main results of this paper. We study how a mean flow with shear modifies the spatial structures of the waves. Our study is on a Gaussian mean flow profile restricted to small latitudes. We show (section~\ref{sec:LowFr}) that especially for large zonal wavenumbers, the spatial structure of the disturbance modes as well as the dispersion relation are both significantly different from the case of zero mean flow. Another objective of the present study is to ask whether the $\beta$-plane system can provide a qualitatively reasonable approximation, at least for a mean flow that is confined very close to the equator. We show, surprisingly, that this is not the case. In fact, the $\beta$-plane approximation shows branches of exponentially growing perturbations whereas there are no instabilities for the full spherical system. Moreover, the disturbance eigenmodes are not confined to small latitudes. Both these results (section~\ref{sec:High_Fr}) bring into question the suitability of the $\beta$-plane approximation.

The layout of the paper is as follows. The basic equations in the $\beta$-plane and the spherical system with relevant approximations, the details of the mean flow whose stability is studied and the numerical method used to get the stability results are presented in section~\ref{sec:Formulation}. In section~\ref{sec:LowFr}, the effects of weak mean shear flow on the propagation and the structure of neutral waves are discussed. Section~\ref{sec:High_Fr} considers the regime where the $\beta$-plane system yields unstable modes. Finally, the results are summarised and their implications are discussed in section~\ref{sec:Summary}.

\section{Problem formulation}\label{sec:Formulation}
\subsection{Governing equations}\label{sec:gov_eq}
The atmosphere and the ocean are predominantly in a state of hydrostatic balance with the spatial scales in the lateral direction being significantly larger than those in the radial direction. The rotating shallow water equations (RSWE) on a sphere, introduced in section~\ref{sec:eqsph}, are thus a natural framework in which a wide range of geophysical phenomena are studied~\citep{vallis_2017}. Depending on the region of interest, these equations could be simplified using a planar approximation, such as the $f$-plane or the $\beta$-plane. The latter, where the Coriolis force is approximated as varying linearly with the latitude, are introduced in section~\ref{sec:eqbeta} and are often used to study equatorial dynamics. The spherical and $\beta$-plane RWSE linearised around a zonally averaged mean flow, as specified in section~\ref{sec:meanflow}, are obtained in section~\ref{sec:eqlin}.

Throughout this paper, $H_0$ indicates the equivalent depth, which provides a link between the horizontal and the vertical dynamics \citep{kiladis2009,paldor2020eqdepth}. While the mean depth of the shallow layer seems like a natural physical choice for the mean height field, the corresponding gravity wave speed is much larger than that observed in the atmosphere. For most of our analysis, we choose $H_0 = 100$~m for our study, which lies between the shallower depth (${\sim}12$--$50$~m) displayed by convectively coupled equatorial waves (CCEW) and a larger depth of ${\sim}200$~m corresponding to the peak projection response of deep convective heating~\citep{WK99}. In \ref{ap:var_H0}, we also show some additional results with different values of $H_0$.

\subsubsection{Equations in the spherical system} \label{sec:eqsph}
In spherical coordinates, RSWE take the following form~\citep{Gill82}:
\begin{subequations}\label{Eq:SWENL_s}
  \begin{align}
    & \frac{\partial u_s^*}{\partial t^*} = 2\Omega \sin\theta v_s^* - \frac{u_s^*}{R\cos\theta}\frac{\partial u_s^*}{\partial \phi}-\frac{v_s^*}{R}\frac{\partial u_s^*}{\partial \theta} + \frac{v_s^* u_s^*\tan\theta}{R}-\frac{g}{R\cos\theta}\frac{\partial h_s^*}{\partial \phi}~,\label{Eq:SWENL_su}\\
    & \frac{\partial v_s^*}{\partial t^*} = -2\Omega \sin\theta u_s^* - \frac{u_s^*}{R\cos\theta}\frac{\partial v_s^*}{\partial \phi} -\frac{v_s^*}{R}\frac{\partial v_s^*}{\partial \theta} - \frac{u_s^{*2}\tan\theta}{R}-\frac{g}{R}\frac{\partial h_s^*}{\partial \theta}~,\label{Eq:SWENL_nv}\\
    & \frac{\partial h_s^*}{\partial t^*} = - \frac{u_s^*}{R\cos\theta}\frac{\partial h_s^*}{\partial \phi}-\frac{v_s^*}{R}\frac{\partial h_s^*}{\partial \theta} - \frac{h_s^*}{R\cos\theta}\left(\frac{\partial u_s^*}{\partial \phi} + \frac{\partial }{\partial \theta}(v_s^*\cos\theta)\right)~.\label{Eq:SWENL_sh}
  \end{align}
\end{subequations}
Here, $u_s^*$ and $v_s^*$ are the dimensional zonal and meridional velocity components respectively and $h_s^*$ is the height of the shallow layer. $\Omega = 2 \pi / 86400\ \textrm{s}^{-1}$ and $R = 6371.22$~km are the angular speed and the radius of the earth respectively, $g = 9.80616$~m/s is the acceleration due to gravity, $\theta\in [-\pi/2, \pi/2]$ is the latitude, and $\phi \in [0, 2\pi]$ is the longitude. Defining the timescale $T_{s} \equiv 1/(2\Omega)$, the non-dimensional variables in spherical coordinates are defined as
\begin{equation}\label{Eq:ND_s}
  (u_s,v_s) = \frac{T_s}{R} (u_s^*,v_s^*)~, \ \ 
  h_s = \frac{h_s^*}{H_0},\ \ t_s = \frac{t^*}{T_s}~.
\end{equation}
The resulting non-dimensional system of equations have the same form as equation~\eqref{Eq:SWENL_s} (formally with $R = 1$, $2 \Omega = 1,$ and $g = 1/\epsilon$) with a single non-dimensional number, the Lamb parameter
\begin{align}
\epsilon \equiv \frac{(2\Omega R)^2}{gH_0}~.
\end{align} 
With $H_0 = 100$~m, note that $\epsilon = 880.44$.
In oceans, the typical equivalent depths are relatively smaller, which results in even larger values of $\epsilon$.

\subsubsection{Equations in the equatorial $\beta$-plane} \label{sec:eqbeta}
The RSWE on the equatorial $\beta$-plane take the following form~\citep{vallis_2017}:
\begin{subequations}\label{Eq:SWENL_b}
  \begin{align}
    &\frac{\partial u_{\beta}^*}{\partial t^*} + u_{\beta}^*\frac{\partial u_{\beta}^*}{\partial x^*} + v_{\beta}^*\frac{\partial u_{\beta}^*}{\partial y^*} - \beta y^* v_{\beta}^* = -g\frac{\partial h_{\beta}^*}{\partial x^*}~, \\
    &\frac{\partial v_{\beta}^*}{\partial t^*} + u_{\beta}^*\frac{\partial v_{\beta}^*}{\partial x^*} + v_{\beta}^*\frac{\partial v_{\beta}^*}{\partial y^*} + \beta y^* u_{\beta}^* = -g\frac{\partial h_{\beta}^*}{\partial y^*}~, \\
    &\frac{\partial h_{\beta}^*}{\partial t^*} + u_{\beta}^*\frac{\partial h_{\beta}^*}{\partial x^*} + v_{\beta}^*\frac{\partial h_{\beta}^*}{\partial y^*} + h_{\beta}^*\left(\frac{\partial u_{\beta}^*}{\partial x^*} + \frac{\partial v_{\beta}^*}{\partial y^*}\right) = 0~,
  \end{align}
\end{subequations}
where $\beta = 2\Omega/R$ is the Coriolis parameter. We define the following reference time and length scales:
\begin{equation}\label{Eq:LT_b}
  T_{\beta} \equiv \beta^{-1/2}(gH_0)^{-1/4}, \ \ 
  L_{\beta} \equiv \beta^{-1/2}(gH_0)^{1/4}~.
\end{equation}
$L_\beta$, known as the Rossby radius of deformation, is the fundamental length scale used in equatorial dynamics and signifies the length scale at which the rotational effects become comparable to the effects due to gravity waves~\citep{vallis_2017}. The gravity wave speed $\sqrt{g H_0}=L_{\beta}/T_{\beta}$ is selected as the reference velocity scale. We can then move to a non-dimensional system with
\begin{equation}\label{Eq:ND_b}
(u_\beta,v_\beta) = \frac{1}{\sqrt{g H_0}}(u_\beta^*,v_\beta^*),\ \
h_\beta = \frac{h_\beta^*}{H_0},\ \ t_\beta=\frac{t^*}{T_\beta}, \ \
x = \frac{x^*}{L_\beta}, \ \ y = \frac{y^*}{L_\beta}~.
\end{equation}
The non-dimensional $\beta$-plane RSWE take the same form as equation~\eqref{Eq:SWENL_b} (formally with $\beta = 1$, $g = 1$) with no non-dimensional parameters appearing in the equations themselves.

\subsection{Base Flow} \label{sec:meanflow}
\begin{figure}
\centering
\includegraphics[width=0.55\textwidth]{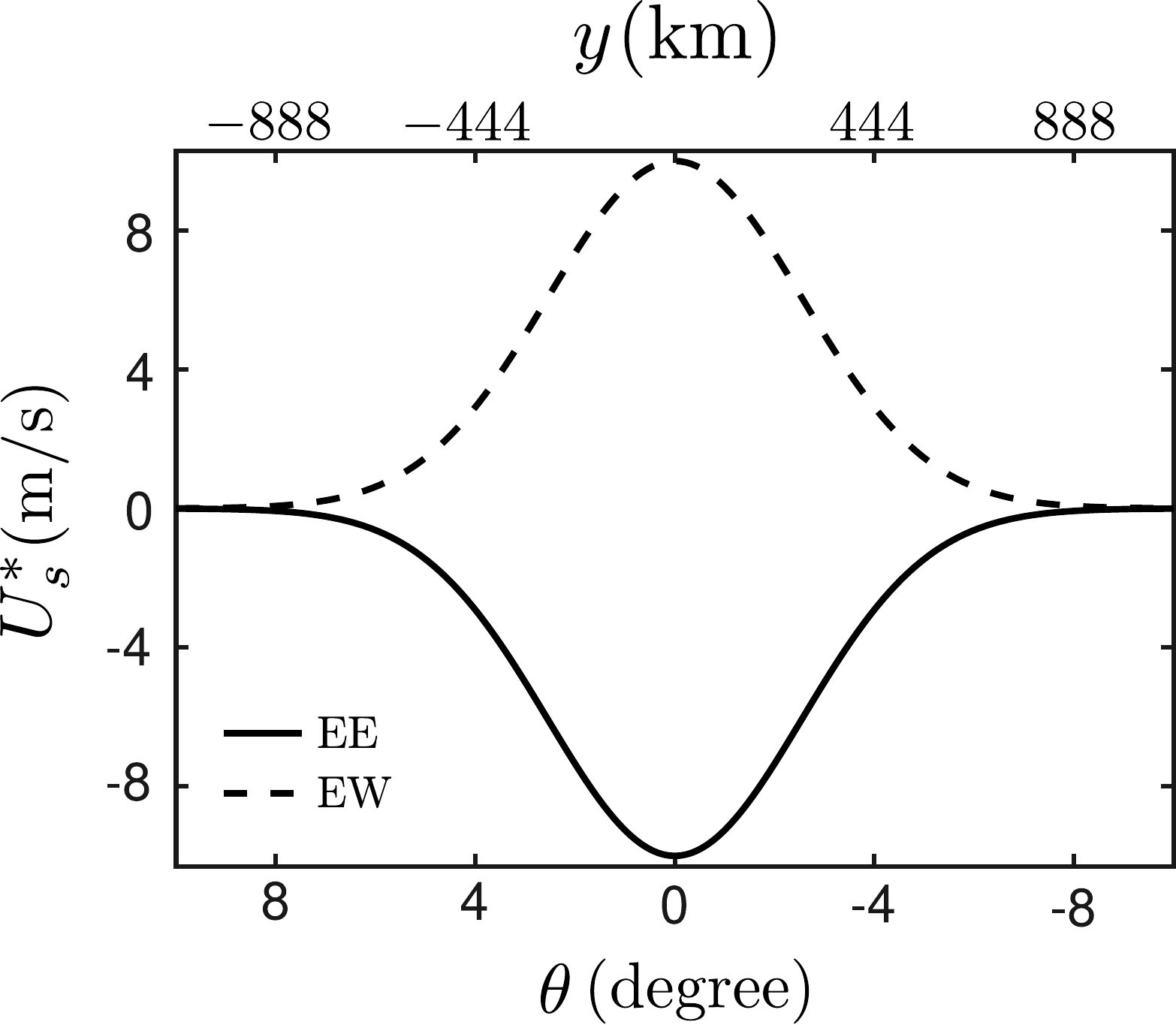}
\caption{Profiles of equatorial easterly (EE) and equatorial westerly (EW) Gaussian mean flow $U_s^{*}(\theta)$, shown here for $\sigma_{\theta} = 0.0628~\textrm{rad}~(3.6^{\circ})$, $\lvert U_0\rvert = 10$~m/s, which corresponds to $\lvert Ro\rvert = 0.011$, i.e., $\lvert Fr\rvert = 0.32$.\label{fig:BF}}
\end{figure}

From climatologically averaged data, it is seen that the zonal mean flow is considerably stronger than the meridional flow~\citep{Plumb2008}. Therefore, a significant number of studies have focused on the dynamics associated with purely zonal mean flows. In this study, we principally consider the stability of equatorial easterly (EE) and equatorial westerly (EW) flows (see figure~\ref{fig:BF}) with Gaussian profiles. To show that the results are of a more generic nature, we also briefly discuss results for a mean flow with a Mexican hat profile in \ref{ap:mex_hat}.

In nondimensional form, the base flow in the spherical system is chosen to be
\begin{equation}
  U_s = Ro \, \exp\left(-\frac{\theta^2}{\sigma_{\theta}^2} \right)~,
\quad V_s = 0~,
\label{eq:usmean} \end{equation}
where $U_s$ and $V_s$ are mean zonal and meridional velocity, respectively with negative or positive $Ro$ being the EE and EW cases respectively. $\sigma_\theta$ sets the latitudinal span of the mean flow. The Rossby number $Ro$ characterises the strength of the mean flow and it is related to the dimensional equatorial speed by 
\begin{equation}
    Ro = \frac{U_0}{2 \Omega R}~.
\end{equation}

For the above flow, the depth of the shallow layer varies $\mathcal{H}_s(\theta) = 1 + H_s(\theta)$ in non-dimensional form. Note that $H_s$ is the variation in the depth of the shallow layer that comes about due to the presence of a mean shear flow. In other words, $H_s = 0$ when $U_s = 0$. $H_s$ is obtained by solving the geostrophic relation in spherical coordinates:
\begin{align}\label{eq:Geost_s}
  \frac{\textrm{d}H_s}{\textrm{d}\theta} = -\epsilon (\sin\theta U_s +  \tan\theta U_s^2)~.
\end{align}

For the $\beta$-plane configuration, the same Gaussian (EE and EW) profiles are used for the mean flow. The distance along a fixed longitude is converted to planar distance in the $\beta$-plane as $y^{*} = R \theta$, which in the non-dimensional form is $y = R \theta / L_\beta$. The required non-dimensional base flow profile is then obtained using the velocity and length scales defined for the $\beta$-plane. Thus we get the mean velocity $U_{\beta} = Fr \, \exp(-y^2/\sigma_y^2)$, where $\sigma_y = R\sigma_{\theta}/L_\beta$ and the Froude number is defined by
\begin{equation}
    Fr = Ro \sqrt{\epsilon} = \frac{U_0}{\sqrt{gH_0}}~.
\end{equation} 
As with the spherical system, the depth of the shallow layer will vary as $\mathcal{H}_\beta(y) = 1 + H_{\beta}(y)$ in the non-dimensional form. Here, $H_\beta$ is the variation of the shallow layer depth when there exists a mean flow ($Fr \neq 0$). $H_\beta$ is obtained by solving
\begin{align}\label{eq:Geost_b}
  \frac{\textrm{d}H_\beta}{\textrm{d}y} = -yU_\beta~.
\end{align}
For the base flow under consideration, $\mathcal{H}_\beta \approx 1$ (or equivalently, $\mathcal{H}_s \approx 1$). \citet{bennett1971,WZ89} considered the variation of $H_\beta$ in the advection term but neglected it in the flux term in the geopotential equation. However, for all of our calculations, we do not make apriori any such assumption and incorporate the variations of the mean equivalent depth due to the background shear flow in all the terms where they appear. 

Throughout this paper, we use $\sigma_y = 400$~km and correspondingly, $\sigma_{\theta} = 0.0628$ \textrm{rad} ($3.6^{\circ}$).

\subsection{Linearised spherical and \texorpdfstring{$\beta$}{beta}-plane systems} \label{sec:eqlin}
For linear stability analysis, we are interested in the asymptotic long-time evolution of perturbations to the base flow. The equations governing the perturbation evolution are rendered linear after neglecting terms that are quadratic in the perturbation quantities. In the spherical system, we invoke an ansatz where the perturbation quantities assume the form $f_s'(\phi,\theta,t_s) = \tilde{f}_s(\theta)\exp(ik\phi -i\omega_st_s)$ with $k$ and $\omega_s$ being the zonal wavenumber and complex frequency respectively. The resulting eigenvalue problem is
\begin{align}\label{Eq:EVP_s}
  &-i\omega_s\tilde{\mathbf{q_s}} = \mathbf{M_s}\tilde{\mathbf{q_s}}~, \\
  \textrm{where } &\tilde{\mathbf{q_s}} = \left[\begin{array}{c} \tilde{u}_s \\ \tilde{v}_s \\ \tilde{h}_s \end{array}\right] \textrm{ and } \mathbf{M}_s = \left[\begin{array}{ccc} M_{uu} & M_{uv}  & M_{uh} \\ M_{vu} & M_{vv} & M_{vh} \\ M_{hu} & M_{hv}& M_{hh} \end{array}\right]\,.
\end{align}
The different operators in $\mathbf{M}_s$ are given below:
\begin{equation}
  \left.
  \begin{aligned}
    &M_{uu} = M_{vv} = M_{hh}=\frac{-ik U_s}{\cos\theta}~,\\
    &M_{uv}=\sin\theta - D_{\theta}U_s+ U_s\tan\theta \equiv \Omega_{as}~, \ \
    M_{uh} = \frac{-ik}{\epsilon\cos\theta}~,\\
    &M_{vu} =-\sin\theta-2U_s\tan\theta~, \ \ 
    M_{vh} = -\frac{1}{\epsilon} D_{\theta}~,\\
    &M_{hu} = \frac{-ik\mathcal{H}_s}{\cos\theta}~, \ \
    M_{hv} = \mathcal{H}_s\left(\tan\theta - D_{\theta}\right) - D_{\theta}\mathcal{H}_s~.    
  \end{aligned}
  \right\}
\end{equation}
In the above, $D_{\theta} = \textrm{d}/\textrm{d}\theta$ and $\Omega_{as}$ is the absolute vorticity of the base flow. We impose homogeneous Dirichlet boundary conditions for the different components of the eigenfunction $\tilde{\mathbf{q_s}}$; $\tilde{u}_s\left(\pm \pi/2\right) = \tilde{v}_s\left(\pm \pi/2\right) = \tilde{h}_s\left(\pm \pi/2\right) = 0$. 

For the $\beta$-plane system, we again consider perturbations of the normal-mode form $f_\beta(x,y,t_\beta) = \tilde{f}_\beta(y)\exp(ik_\beta x -i\omega_\beta t_\beta)$ with $k_\beta$ and $\omega_\beta$ being the non-dimensional zonal wavenumber and complex frequency respectively. With this ansatz, the resulting  eigenvalue problem is:
\begin{align}
  &-i\omega_\beta\tilde{\mathbf{q}}_\beta = \mathbf{M}_\beta\tilde{\mathbf{q}}_\beta~, \label{Eq:EVP_b}\\
  \textrm{where } &\tilde{\mathbf{q}}_{\beta} = \left[\begin{array}{c} \tilde{u}_{\beta} \\ \tilde{v}_{\beta} \\ \tilde{h}_{\beta} \end{array}\right] \textrm{ and } \mathbf{M}_\beta = \left[\begin{array}{ccc} -ik_\beta U_\beta & \Omega_{a\beta} & -ik_\beta \\ -y & -ik_\beta U_\beta & -D_y \\ -ik_{\beta}\mathcal{H}_{\beta} & yU_{\beta} - \mathcal{H}_{\beta} D_y & -ik_\beta U_\beta \end{array}\right]~.
\end{align}
In the above, $D_y = \textrm{d}/\textrm{d}y$ and $\Omega_{a\beta} (\equiv y - D_y U_\beta)$ is the absolute vorticity of the base state. For boundary conditions, the perturbation fields are specified to go to zero as $y \rightarrow \pm \infty$.

In both the above cases, a modal (exponentially growing) instability exists when there is at least one eigenvalue with a positive imaginary part, i.e., when $\omega_s$ or $\omega_\beta$ has a positive imaginary part.

We also note that the wavenumbers in the spherical system $k$ and in the $\beta$-plane system $k_\beta$ are related by $k = R\cos\theta_0 k_\beta/L_\beta$, where $\theta_0$ is the reference latitude about which the $\beta$-plane is defined. In the present study, as our $\beta$-plane is centred about the equator, we have $\theta_0 = 0$. This then gives us $k_{\beta} = L_\beta k/R$. 

\subsection{Numerical method}
Equations~\eqref{Eq:EVP_s} and~\eqref{Eq:EVP_b} are solved numerically using a Chebyshev collocation method. The general method has been extensively validated and used in several contexts (see e.g.~\cite{ JosePRF,Croor20}). Our starting point is the set of $N+1$ Chebyshev collocation points $\xi_j \in [-1,1]$: 
\begin{equation}\label{eq:Cheb}
  \xi_j= \cos \frac{j\pi}{N},~j=0,~1,~\cdots,~(N-1),~N.
\end{equation}
Differentiation matrices, which give derivatives of functions defined on these collocation points in discrete form are well-known~\citep{canuto2007,Trefethen}. The Chebyshev grid $\xi\in[-1,1]$ is mapped appropriately to the domains of interest in the $\beta$-plane $y\in(-\infty,\infty)$ and in spherical coordinates $\theta\in\left[-\pi/2,\pi/2\right]$, in such a way that the grid is clustered to provide a higher density of grid points in regions of large variations.

In the $\beta$-plane, the algebraic mapping
\begin{equation}\label{eq:Map_b}
  y=\frac{\alpha\xi}{\sqrt{1-\xi^2}}.
\end{equation}
is used to transform the domain of interest ($y$) to the Chebyshev grid ($\xi$). In general logarithmic or exponential maps could also be used~\citep{Boyd01}.  Under this transformation, the derivative with respect to $y$ is given as:
\begin{equation}
  \frac{\textrm{d}}{\textrm{d}y}=\frac{(1-\xi^2)^{3/2}}{\alpha}\frac{\textrm{d}}{\textrm{d}\xi}.
\end{equation}
We use the above to pre-multiply the Chebyshev differentiation matrix appropriately to obtain the discrete differentiation matrix in $y$. For every value of $\alpha$, the points $\xi = \pm 1$ are mapped to $y \rightarrow \pm \infty$. For low values of $\alpha$ (${\sim}0.1$), the number of points in $y$ are insufficient in the far-field. On the other hand, for larger values of $\alpha$ (${\sim}20$), the grid is poorly resolved in the vicinity of the equator. For all our calculations, we use $\alpha = 1$; we have verified that the results do not change when $\alpha$ is changed to 0.8 or 2.

For the spherical system, the Chebyshev grid ($\xi$) is first stretched and then scaled by a factor of $\pi/2$, such that the two grids are related as~\citep{vinokur1983,Rama04}:
\begin{align}\label{eq:Map_s}
  &\theta = \frac{a\pi}{\sinh\left(by_0\right)}\left[\sinh b\left(\frac{\xi+1}{2}-y_0\right)+\sinh \left(by_0\right) \right]- \frac{\pi}{2}, \\
  \textrm{where } &y_0=\frac{1}{2b}\log\frac{1+a\left(\textrm{e}^{b}-1\right)}{1+a\left(\textrm{e}^{-b}-1\right)}.
\end{align}
In the above, $\pi(2a-1)/2$ is the location where the grid points are concentrated and $b$ is a stretching parameter. Given that our interest primarily lies in the region about the equator, we select $a=0.5$. In this study, we have varied $b$ between 4 and 8 and ensured that the results do not change. We now have
\begin{align}
  \frac{\textrm{d}}{\textrm{d}\theta}= \frac{2}{\pi ab}\sinh \left(by_0\right) \sech b\left(\frac{\xi+1}{2}-y_0\right)\frac{\textrm{d}}{\textrm{d}\xi.}
\end{align}

For both the systems, we select $N=200$. We have verified that the eigenvalues match to ten significant digits upon changing $N$ to 300. We validate our numerical approach for the $\beta$-plane by reproducing the spectrum of the Rossby modes ~\cite[Figure 14]{WZ89}. As a validation of our numerical approach in spherical coordinates, the growth rates as a function of $H_0$ for $k = 1$ as reported in ~\cite{paldor2021} (Figure 5 therein) were recovered.

\section{Low Froude number regime: neutral waves}\label{sec:LowFr}
Under low levels of shear, both the $\beta$-plane and spherical systems yield spectra consisting entirely of neutral modes. We demonstrate our findings on $\lvert U_0\rvert = 0.5$~m/s, which corresponds to
$\lvert Fr\rvert = 0.02$.

\subsection{Spectra}\label{ssec:Spectra}
The configuration in the $\beta$-plane setting with no mean flow ($U_{\beta} = 0$) has been studied extensively~\citep{holton2004,vallis_2017}, and the non-dimensional eigenfrequencies are given by
the dispersion relation~\citep{Matsuno}
\begin{equation}\label{eq:wk_U0_b}
  \omega_\beta^2-k_\beta^2-\frac{k_\beta}{\omega_\beta}  = n+\frac{1}{2}, \quad n = 0, 1, 2, \dots.
\end{equation}
In the above, $n$ is the meridional mode number, which gives the number of zeros in the meridional velocity field along a given longitude. The cubic equation~\eqref{eq:wk_U0_b} yields three families of waves: EIG, WIG, and Rossby waves for $n \ge 1$. For $n = 0$, the WIG solution is unphysical and the Rossby wave solution is commonly known as the MRG mode. In addition to these modes, there exists the Kelvin mode with $\omega_{\beta}=k_{\beta}$ for $n = -1$.

On introducing mean shear, we find that the MRG and Rossby modes are significantly affected. In contrast, the effects of mean shear on the EIG, WIG and Kelvin modes are not as prominent. This is consistent with results from earlier studies (e.g. see~\cite{WZ89}). Therefore, the subsequent discussion will focus on MRG and Rossby waves with the corresponding $\beta$-plane theory for zero mean flow case used as a template for comparison.

\begin{figure}
  \includegraphics[width=\textwidth]{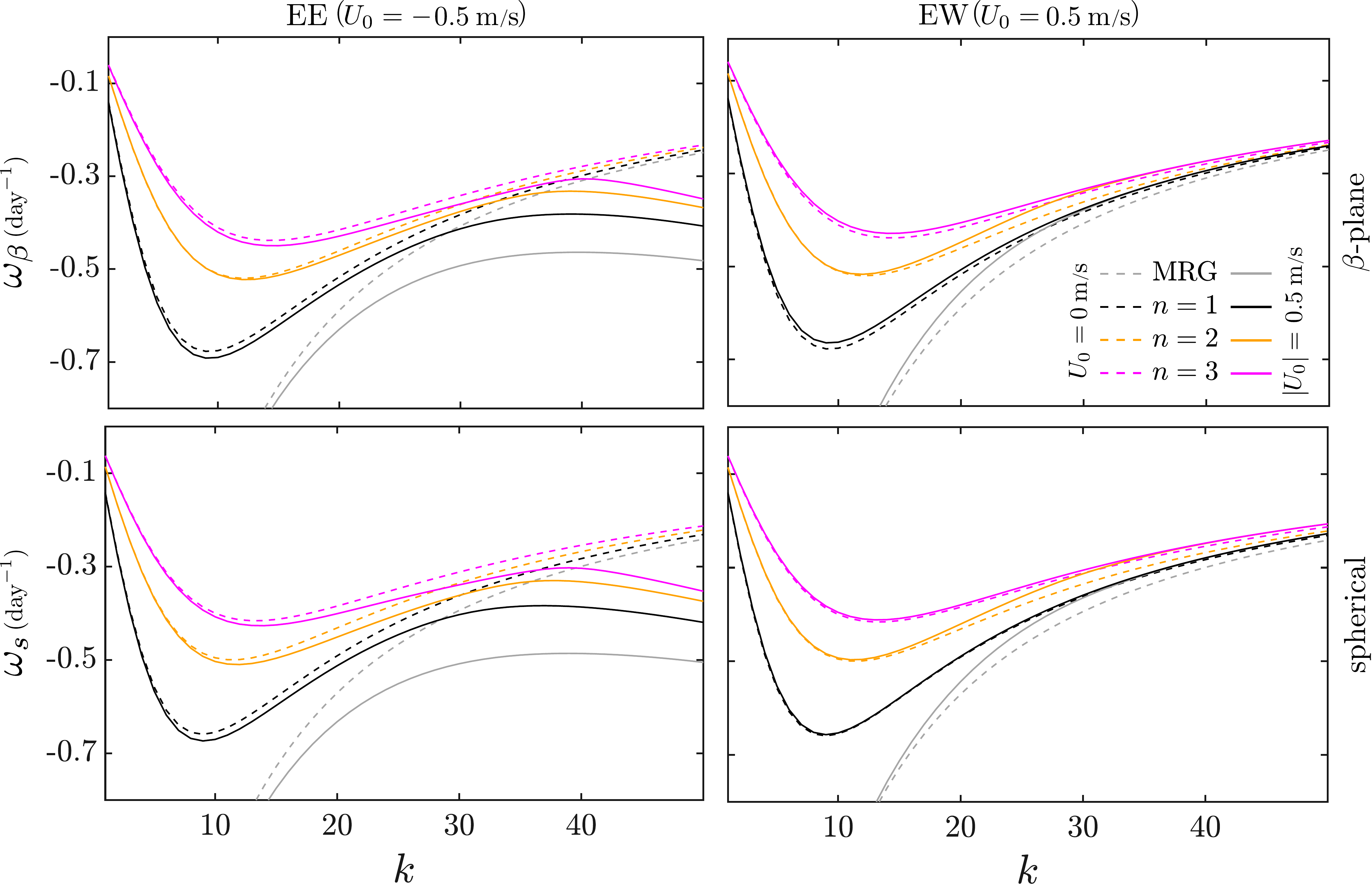}
  \caption{Dispersion curves for the Rossby and MRG waves in the $\beta$-plane (top) and the spherical system (bottom) for EE (left) and EW (right). Solid lines correspond to $\lvert U_0\rvert=0.5$~m/s ($\lvert Fr\rvert=0.02$) and dashed lines are for $U_0 = 0$ (no mean flow). For equatorial westerly flow (EW), the frequencies corresponding to successive $n$ are closely clumped for large $k$. For equatorial easterly flow (EE), the short waves (large $k$) tend to be non-dispersive, in contrast with the $U_0 = 0$ short waves that are dispersive with $\omega \sim -k^{-1}$. \label{fig:wr_k_low_Fr}}
\end{figure}

Figure~\ref{fig:wr_k_low_Fr} shows the dispersion curves of MRG and Rossby modes for the first few values of $n$. The waves travel westward with a higher phase speed in EE than in EW, a characteristic also noted in~\cite{WZ89}. From the figure, for both EE and EW, it is apparent that the deviations in the frequency in the presence of shear flow are more pronounced for short waves (high $k$). When the mean shear flow is weak, we see that the spectra are qualitatively similar in the spherical system and the $\beta$-plane. The eigenstructure however has pronounced differences, and we will return to this point.

MRG and Rossby waves have much slower phase speeds than the other families of waves. In the absence of shear, the dispersion relation for Rossby waves can be approximated by neglecting $\omega_\beta^2$ in equation~\eqref{eq:wk_U0_b}~\citep[section 8.2]{vallis_2017} to get:
\begin{align}
  \omega_\beta \sim -\frac{k_\beta}{2n+1+k_\beta^2}.
\end{align}
The long waves ($k_\beta\rightarrow 0$) turn out to be effectively non-dispersive, propagating with a constant phase speed of $\omega_\beta/k_\beta\sim-(2n+1)^{-1}$. On the other hand, short waves ($k_\beta\rightarrow \infty$) are dispersive with the frequency scaling as $\omega_\beta \sim -k_\beta ^{-1}$. It can also be noted that the frequencies of MRG and Rossby modes remain distinct for sufficiently large values of $k_\beta $ in the absence of shear flow.

With EE, long waves continue to remain non-dispersive as in the zero mean shear configuration. In contrast to the case with no shear, however, the dispersion curve turns around, and short waves again tend to become non-dispersive (showing a constant phase speed) for each $n$ that is lower than the Kelvin wave phase speed. With EW, waves of successive $n$ display frequencies which clump closer to each other with increasing $k$. This clumping is less in the spherical system for a given $k$. Using the base flow profiles considered by \cite{WZ89}, we were able to discern similar characteristics for short waves (not shown). In their study, these properties were not as evident owing to the smaller range of wavenumbers considered.

\begin{figure}
  \includegraphics[width=1\textwidth]{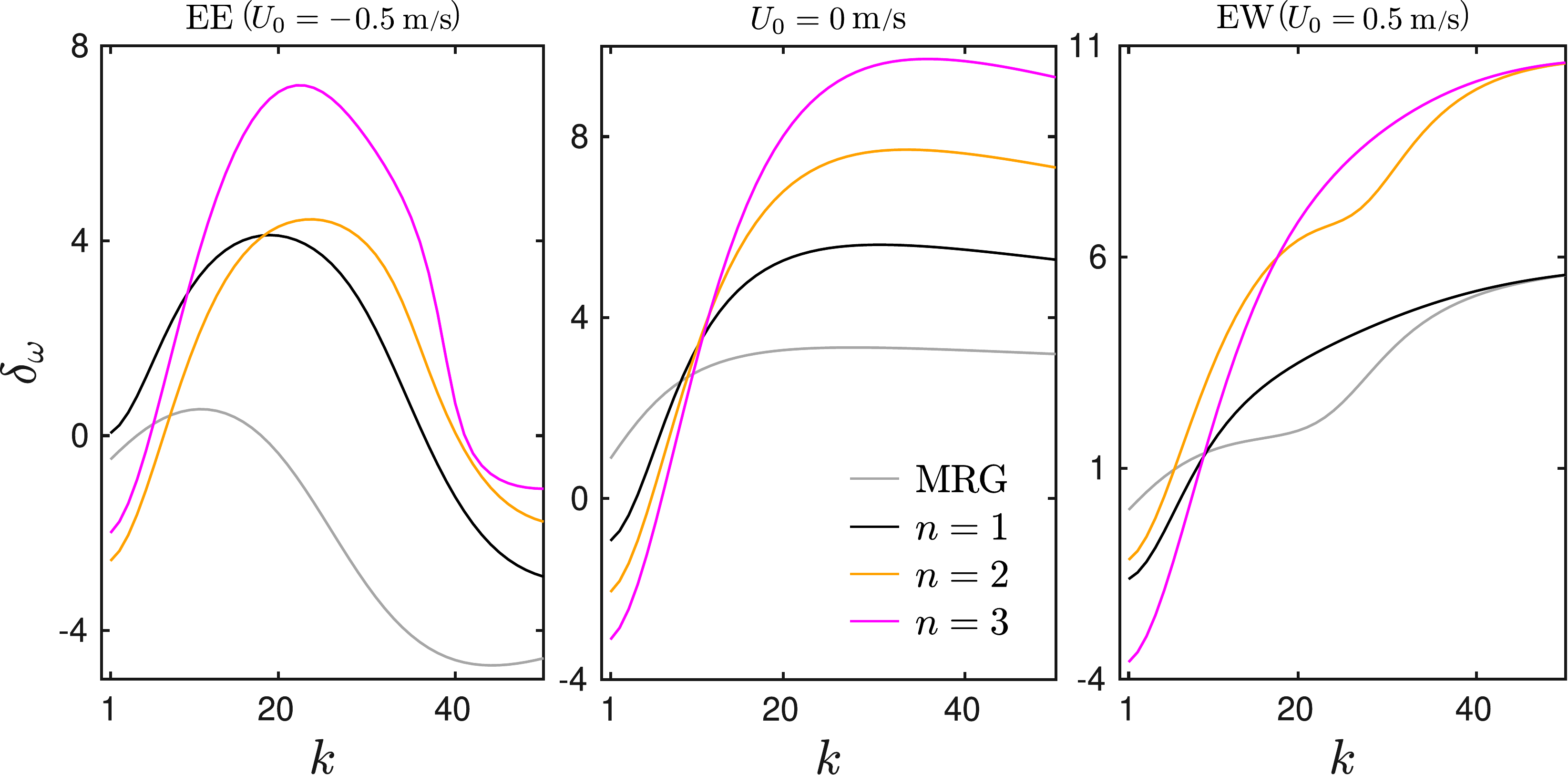}
  \caption{Relative difference $\delta_\omega$ (defined in equation~\eqref{eq:reldiff}) between spectra of the $\beta$-plane and the spherical systems for MRG ($n = 0$) and Rossby modes ($n \ge 1$) for EE (left), no mean flow (centre), and EW (right). Even for a base flow confined to low latitudes, the deviations between the spectrum of the $\beta$-plane and spherical system are significant at moderate to high $k$, especially for EW.\label{fig:dwr_k}}
\end{figure}

In order to compare the spectra of the $\beta$-plane and spherical systems, we compute the percentage difference as
\begin{align}
  \delta_\omega = \frac{\omega_\beta-\omega_{s}}{\omega_{s}}\times 100,\label{eq:reldiff}
\end{align}
and plot it in figure~\ref{fig:dwr_k} for the case of no mean shear (middle), EE (left) and EW (right). In the absence of shear, there is already a deviation between results from the two systems, but this remains bounded within $10\%$. The frequency obtained under the $\beta$-plane approximation is seen to be underestimated for a majority of the long waves (small $k$). For short waves, there is a larger overestimation of the phase speeds, e.g., in the $n=2$ and $n=3$ Rossby modes with EW as the mean flow. It is curious that the margin of error is low for smaller $k$ regardless of the type of mean flow, whereas intuition would not suggest this. In the following subsection, we examine the corresponding eigenfunctions to fully appreciate the limitations of the $\beta$-plane approximation.

\subsection{Eigenfunctions}
In the $\beta$-plane setting in the absence of shear flow, the meridional velocity takes the form~\citep{Matsuno}:
\begin{equation}\label{eq:EFv_U0_b}
  v_\beta(y) = He_n(y)\exp(-y^2/2),
\end{equation}
where $He_n(y)$ is the $n$th order Hermite polynomial having $n$ zeros \citep{Swarttouw2010}. As was done for the analysis of the spectra, we continue to employ the meridional mode number $n$ to classify the eigenfunctions.

\begin{figure}
  \includegraphics[width=1.15\textwidth]{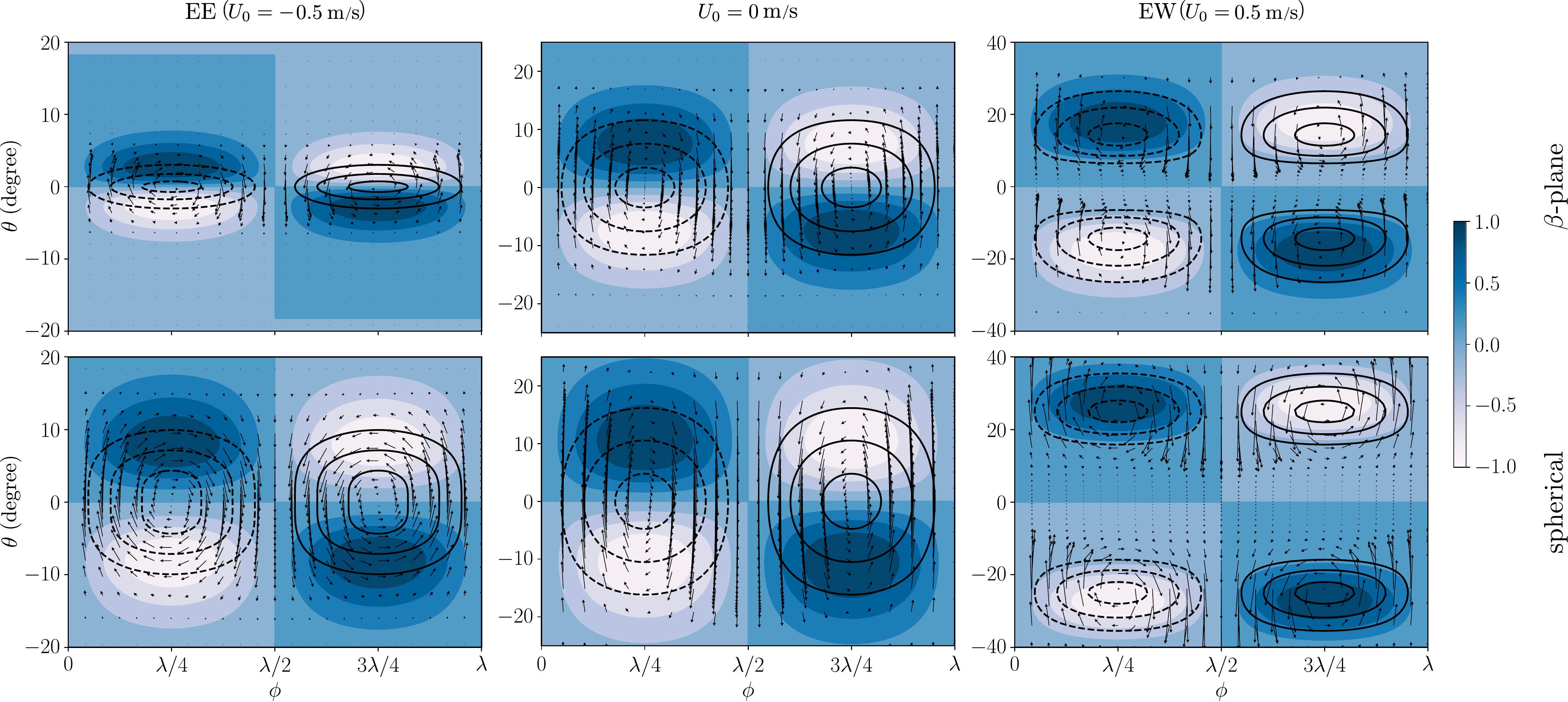}
  \caption{The MRG ($n=0$) modes for $k=50$ in the $\beta$-plane (top) and spherical (bottom) systems with EE (left), no shear (centre) and EW (right); the perturbation wavelength is $\lambda = 2\pi/k = \pi/25$ here. Line contours and arrows depict the vorticity and the velocity respectively; colour contours depict the surface elevation. The dashed and continuous contours correspond, respectively, to the negative and positive values of vorticity. Note that the limits in the latitude are not the same across the different panels. With shear, eigenfunctions are clearly less confined in the spherical system than in the $\beta$-plane, and in EW than EE.}
  \label{fig:EF_n_0_k_50}
\end{figure}

\begin{figure}
  \includegraphics[width=1.15\textwidth]{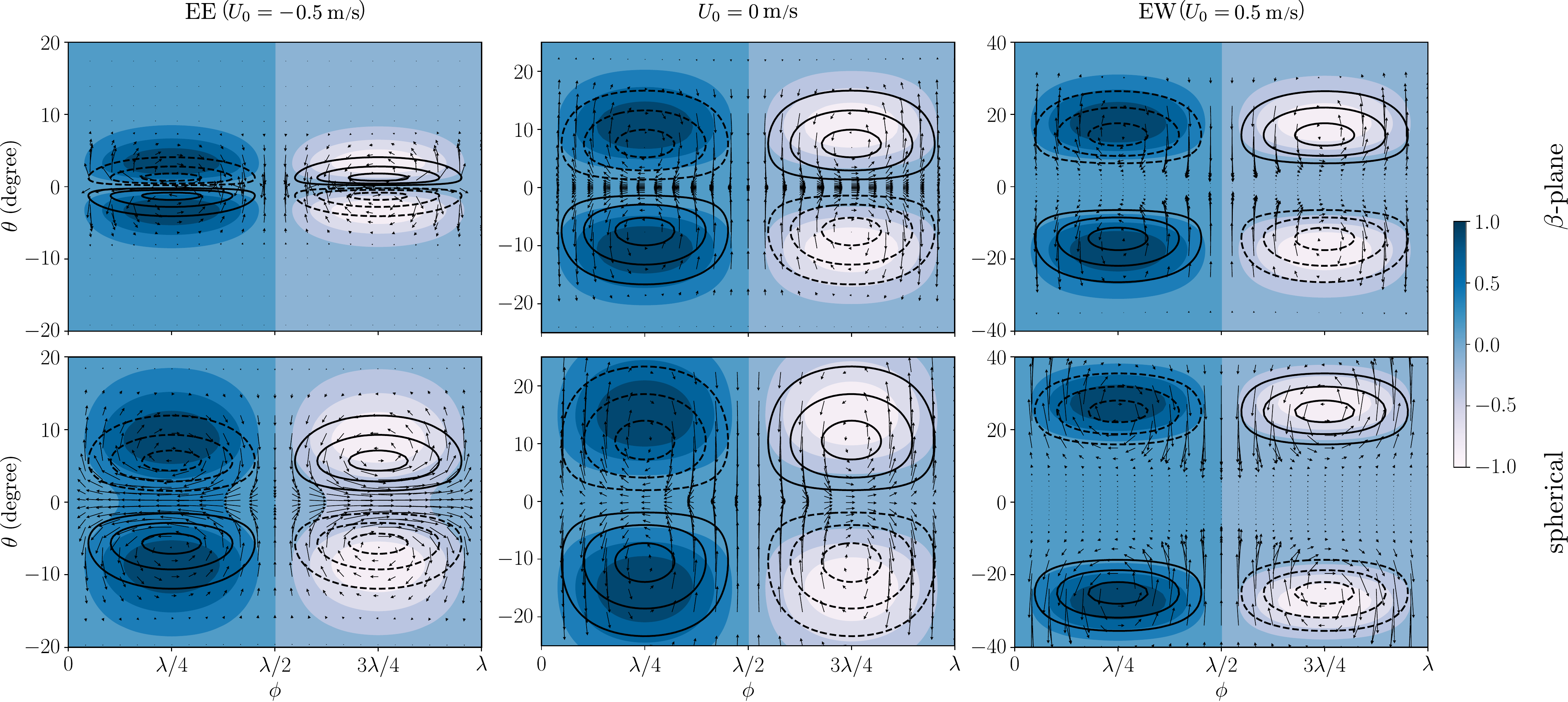}
  \caption{Same as figure~\ref{fig:EF_n_0_k_50}, but for $n=1$. Again, the eigenfunctions are more confined to the equatorial zone in the $\beta$-plane than on a sphere in the presence of shear.}
  \label{fig:EF_n_1_k_50}
\end{figure}

In figures~\ref{fig:EF_n_0_k_50} and~\ref{fig:EF_n_1_k_50}, the $n = 0$ MRG and $n = 1$ Rossby modes are shown in the different mean flow configurations for the $\beta$-plane and spherical settings. We choose $k = 50$ for a demonstration of the findings, which corresponds to a wavelength of about $800$ km. This example is characteristic of the entire range of high wavenumbers. Note that for plotting eigenfunctions obtained in the $\beta$-plane setting, we use $\theta = L_{\beta}y/R$. In the absence of mean flow, a primary feature of modes with even $n$ is a longitudinal cross equatorial flow with a vortex cell centered at the equator. While this feature is still apparent for EE, the mode is characterised by two off-equatorial vortices accompanied by a quiet band around the equator when the mean flow is EW.  When $n$ is odd, with zero mean flow, the eigenfunction is characterised by off-equatorial vortices with a strong equatorial zonal flow. Once again, with EE as the mean flow, these qualitative features are retained. With EW, as before, the mode is quiescent near the equator with prominent off-equatorial vortices. With regard to the surface elevation, the fields are symmetric and antisymmetric about the equator when $n$ is odd and even respectively.

Whether in EE or EW, the most significant effect of shear is that the eigenfunction is far more spread out in the meridional direction for the complete spherical system, and the $\beta$-plane approximation misses this. Significantly, the disturbance flow extends into the extratropics, far beyond the mean flow, which is negligible beyond 8 degree north or south. The visual evidence of figures~\ref{fig:EF_n_0_k_50} and~\ref{fig:EF_n_1_k_50} can be made more quantitative. To this end, for an eigenfunction $\tilde{\mathbf{q}}$, we define:
\begin{align}
  &\lVert \tilde{\mathbf{q}}\rVert^2(\theta) = \frac{1}{4}\int^{\theta}_{-\theta} \textrm{d}\theta'~\cos\theta' \left(\tilde{u}^{\dagger}\tilde{u} + \tilde{v}^{\dagger}\tilde{v} + \tilde{h}^{\dagger}\tilde{h} \right)~, \label{eq:L2_meas_q} \\
  \textrm{and }&\tau(\theta) = \frac{\lVert \tilde{\mathbf{q}}\rVert^2(\theta)}{\lVert \tilde{\mathbf{q}}\rVert^2(\pi/2)} \quad \textrm{ for } 0 \leq \theta \leq \frac{\pi}{2}~. \label{eq:tau_q}
\end{align}
In the above, the dagger symbol ($\dagger$) represents the complex conjugate. Note that $\lVert \tilde{\mathbf{q}}\rVert^2$ (and hence $\tau$) monotonically increases with $\theta$. As a result, we can define a threshold latitude $\theta_\tau$ as the value of $\theta$ for which $\tau$ takes a specific value; we choose $\tau = 0.9$ for our analysis. A small (large) $\theta_\tau$ is associated with a high (low) degree of equatorial confinement.

\begin{figure}
  \includegraphics[width=\textwidth]{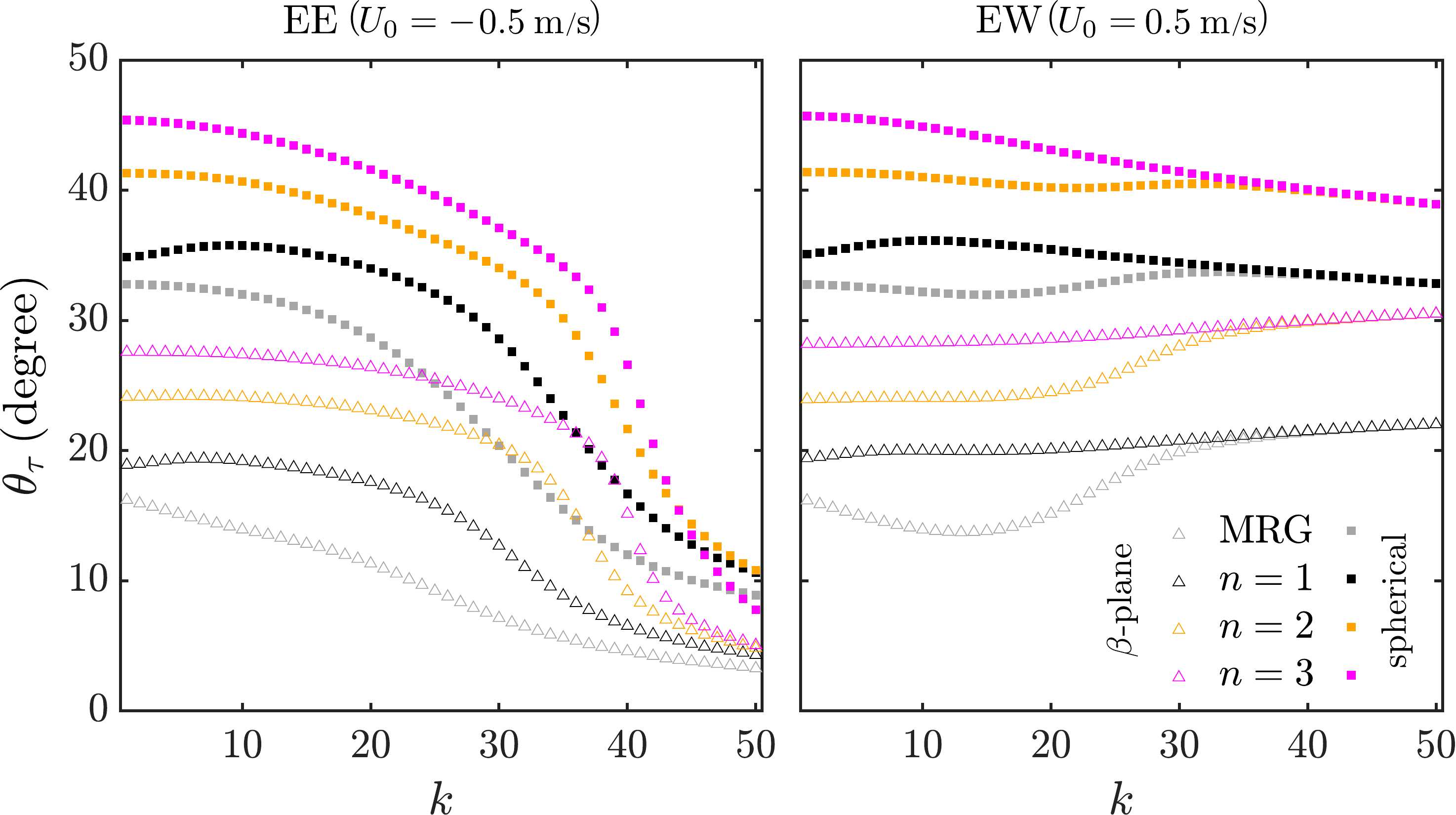}
  \caption{The latitude $\theta_\tau$ (for $\tau = 0.90$) as a function of the wavenumber $k$ in the $\beta$-plane (triangles) and the spherical system (squares) for EE (left) and EW (right). The eigenfunctions are highly equatorially confined in the $\beta$-plane system compared to the spherical system for all the families of waves across all wavenumbers in both EE and EW. \label{fig:Thet_cr}}
\end{figure}

Figure~\ref{fig:Thet_cr} shows $\theta_\tau$ as a function of $k$ for MRG and Rossby waves in EE and EW. It is immediately evident that the findings of figures~\ref{fig:EF_n_0_k_50} and~\ref{fig:EF_n_1_k_50} are applicable across the wavenumber range for all wave families. Even though the mean flow is extremely confined equatorially, the perturbation waves it creates extend poleward well into the extratropics. This important physics is missed by making a $\beta$-plane approximation. This approximation causes the disturbance energy to become highly trapped equatorially. Consistent with the zero mean flow scenario, for lower values of $k$, the degree of confinement reduces as $n$ increases. The short waves (large $k$) are significantly more equatorially trapped for EE than EW, showing the same trend as results pertaining to the meridional scale of Rossby waves for base flow of \cite{WZ89}. For EW at higher $k$ it was seen (right panels of figure~\ref{fig:wr_k_low_Fr}) that eigenvalues belonging to successive $n$ were clustered together (see section~\ref{ssec:Spectra}). We see a similar feature in the degree of the confinement of the eigenfunctions in figure~\ref{fig:Thet_cr}. For larger $k$, $\theta_{\tau}$ get closer for successive values of $n$.

For insight into the localisation features of the modes, we utilise arguments based on potential vorticity (PV), which is a materially conserved quantity in shallow water systems. In the presence of a mean flow varying with latitude, in addition to the planetary vorticity, the gradient of the mean flow alters the ambient PV. Denoting the ambient PV as $Q$ (with appropriate subscripts), the expressions in the two settings, take the form:
\begin{align}
  &Q_{s} = \frac{1}{1 + H_s}\left(\sin\theta - \frac{\textrm{d}U_s}{\textrm{d}\theta} + U_s\tan\theta \right)~, \label{eq:APV_sph} \\
  \textrm{and } &Q_{\beta} = \frac{1}{1 + H_{\beta}}\left(y - \frac{\textrm{d}U_{\beta}}{\textrm{d}y}\right)~.\label{eq:APV_beta}
\end{align}
The meridional gradient of the ambient PV plays a crucial role in propagation the Rossby modes \citep[section 3.16]{Pedlosky_1987}; we refer to this gradient as $DQ$. As a fluid parcel moves to a different latitude, its associated relative vorticity gets modified so as to conserve the total PV at the displaced location. This induced relative vorticity acts as a restoring force of the parcel. Greater $DQ$ implies a greater restoring force experienced by the parcel, which in turn leads to more trapping in the vicinity of the equator. For the mean flows considered here, we note that the $DQ$ in the $\beta$-plane is greater than that in the spherical system. Hence, we expect the eigenfunctions in the spherical system to have larger meridional extents.

Next, we examine the characteristics of the perturbation zonal velocity and surface elevation. For a given eigenmode, these quantities can be expressed in terms of the perturbation meridional velocity and its meridional gradient as follows:
\begin{align}
&\tilde{u}_{s} = i\mathcal{S}_{uv}\tilde{v}_{s} + i\mathcal{S}_{uDv}D_y\tilde{v}_{s},~\tilde{h}_{s} = i\mathcal{S}_{hv}\tilde{v}_{s} + i\mathcal{S}_{hDv}D_y\tilde{v}_{s}~,\label{eq:uhs_vs} \\
&  \tilde{u}_{\beta} = i\mathcal{B}_{uv}\tilde{v}_{\beta} + i\mathcal{B}_{uDv}D_y\tilde{v}_{\beta},~\tilde{h}_{\beta} = i\mathcal{B}_{hv}\tilde{v}_{\beta} + i\mathcal{B}_{hDv}D_y\tilde{v}_{\beta}~. \label{eq:uhb_vb}
\end{align}
\ref{ap:uh_v} gives explicit formulae for the coefficients above as functions of the base flow state and the eigenvalue. We define the number of zeros of the zonal velocity and the surface elevation as $n_u$ and $n_h$ respectively. For the zero mean flow case in the $\beta$-plane setting, with $\tilde{v}_{\beta}$ given by equation~\eqref{eq:EFv_U0_b}, $n_u = n_h = n + 1$ for the MRG and Rossby modes. 
We report that this relationship for $n_h$ continues to hold in the presence of shear flow for the range of wavenumbers considered here although there are notable changes in $n_u$, as discussed below. 

\begin{figure}
  \includegraphics[width=1\textwidth]{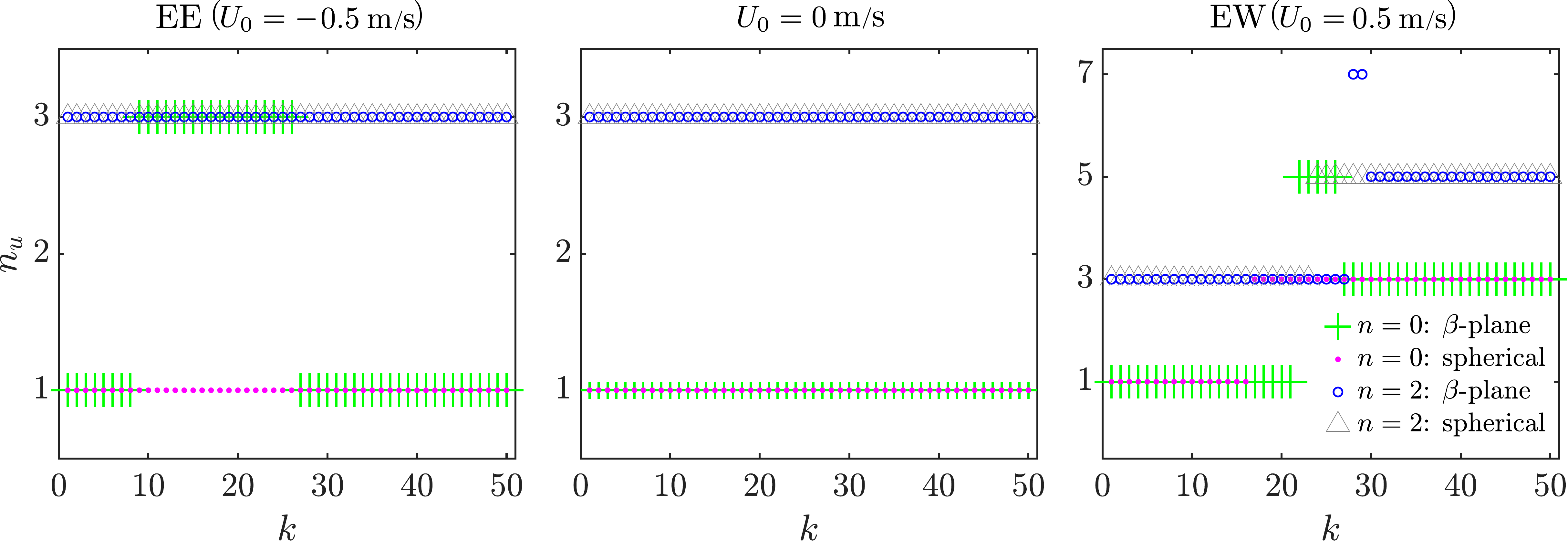}
  \caption{Number of zeros ($n_u$) in latitudinal variation of the zonal perturbation velocity field ($\tilde{u}$) as a function of $k$ in the presence of EE (left), no shear (centre) and EW (right) in the $\beta$-plane and spherical systems. Compared to the zero mean shear case, the relationship $n_u=n+1$ does not hold in the presence of shear for a range of values of $k$. This leads to the bifurcation of the equatorial vortex of eigenfunctions corresponding to even $n$ into two off-equatorial vortices (see section~\ref{sec:LowFr}). \label{fig:nu_k}}
\end{figure}

In the zero mean flow case, we have $n_u = n + 1$ for all $k$. The corresponding velocity field involves either a cross-equatorial flow (even $n$) or an equatorial zonal flow (odd $n$).  We report that when $n$ is odd, $n_u$ remains unchanged for the non-zero mean flow cases. In contrast, for even $n$, there are notable changes in $n_u$ with change in the wavenumber. In figure~\ref{fig:nu_k}, we plot $n_u$ as a function of the wavenumber for $n = 0$ and $n = 2$. The central panel shows the zero mean flow case to serve as a reference, and the left and right panels are for EE and EW with $\lvert U_0\rvert = 0.5$~m/s. With EE as the mean flow, the perturbation velocity characteristics around the equator are similar to those obtained in the zero mean flow case. There is a range of wavenumbers for which $n_u = n + 3$ in the $\beta$-plane setting (see the left panel of figure~\ref{fig:nu_k}).

The differences in $n_u$ are more stark when EW is the mean flow. For lower values of $k$, the cross-equatorial flow continues to be prominent. As $k$ increases, the eigenfunction structure comprises two off-equatorial vortices in addition to the equatorial vortex. These off-equatorial vortices become more pronounced with increasing $k$. While $n_u = n + 1$ for the large part, there are some values of $k$ for which $n_u = n + 5$ in the $\beta$-plane setting. As $k$ increases further, the off-equatorial vortex cells become dominant and the flow in the vicinity of equator is considerably weaker. Recall the nearly quiescent regions around the equator in figures~\ref{fig:EF_n_0_k_50} and~\ref{fig:EF_n_1_k_50}. In this range of $k$, we have $n_u = n + 3$.

\begin{figure}
  \includegraphics[width=\textwidth]{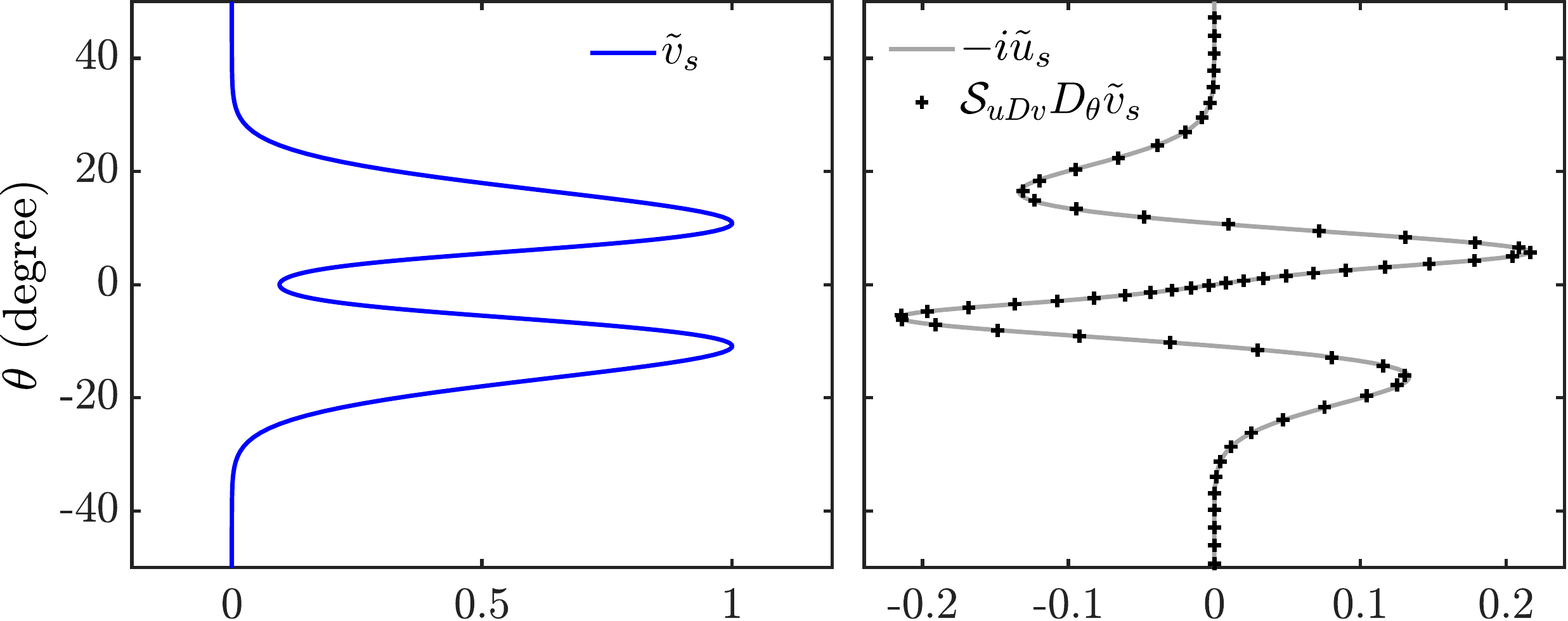}
  \caption{The velocity components of the MRG mode (in spherical coordinates) when $k = 40$ and $U_0 = 0.5$~m/s (EW). In the left panel, the meridional velocity $\tilde{v}_s$ is seen to have three local extremas. On the right, the dominant contribution to the zonal velocity $\tilde{u}_s$ is shown to be from the term proportional to $D_{\theta}\tilde{v}_s$. For this mode, $n = 0$ and $n_u = 3$.}
  \label{fig:nu_arg}
\end{figure}

Using equations \ref{eq:uhs_vs} and \ref{eq:uhb_vb}, we can understand the changes in $n_u$ by examining the meridional velocity $\tilde{v}$ and its meridional gradient $D\tilde{v}$; as the arguments hold in both settings, we drop the subscripts that distinguish the quantities in the $\beta$-plane and the spherical system. We find that the dominant contribution to the zonal velocity $\tilde{u}$ is due to the term proportional to $D\tilde{v}$, and consequently $n_u$ is precisely the number of sign changes in $D\tilde{v}$. We show this explicitly for the MRG mode (with $n = 0$) when $k = 40$ and $U_0 = 0.5$~m/s in figure \ref{fig:nu_arg}. Therefore, the seemingly abrupt jumps in $n_u$ are accounted for by the changes in the number of local extrema of the meridional velocity $\tilde{v}$.
\section{High Froude number regime}\label{sec:High_Fr}
We now turn our attention to the stability characteristics of base flow configurations with larger $\lvert U_0\rvert$. Although the mean flow profile considered is idealised, our choices for $\lvert U_0\rvert$ here are motivated by the mean velocities at the equator obtained from climatological data~\citep{Plumb2008}. We discuss four cases with $\lvert U_0\rvert = 5, 10, 15, 20$~m/s, which correspond to $Fr = 0.159, 0.319, 0.479, 0.639$ and $Ro = 0.0054, 0.0108, 0.0161, 0.0215$.

\begin{figure}
  \includegraphics[width=\textwidth]{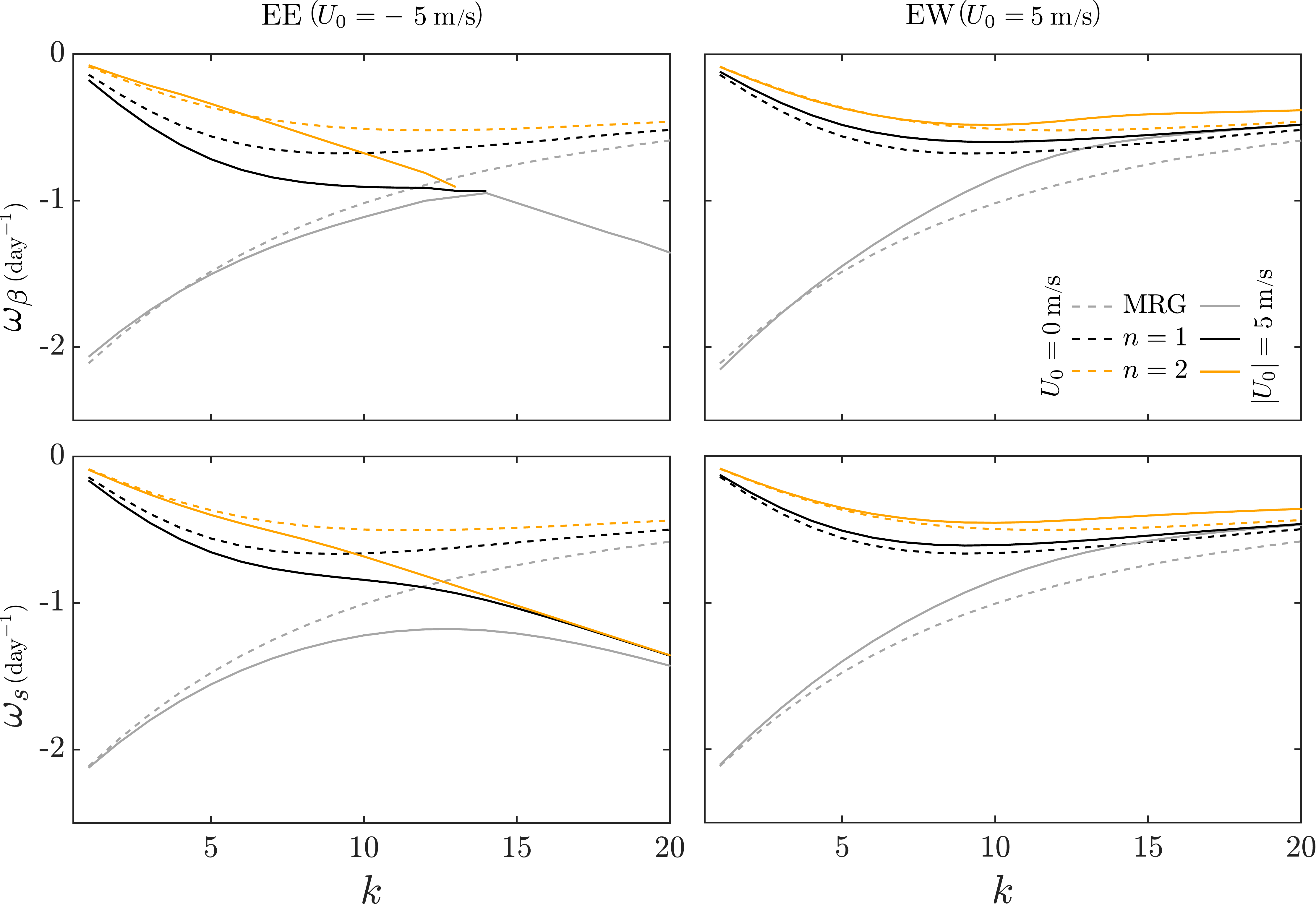}
  \caption{Dispersion curves for the Rossby and MRG waves in the $\beta$-plane (top) and the spherical system (bottom) for EE (left) and EW (right). Solid lines correspond to $\lvert U_0\rvert=5$~m/s and dashed lines are for $U_0 = 0$ (no mean flow). For equatorial westerly flow (EW), the frequencies corresponding to successive $n$ are closely clumped for large $k$. For equatorial easterly flow (EE), the short waves (large $k$) tend to be non-dispersive. \label{fig:wr_k_high_Fr}}
\end{figure}

When compared with the low shear cases, the classification of modes becomes relatively challenging. As in the low shear configurations, the modes are categorised into different wave families based on the number of zeros of its meridional velocity component.   
We first examine the changes in the features of the MRG and Rossby modes with increasing shear. Figure \ref{fig:wr_k_high_Fr} shows the eigenfrequency of MRG and Rossby for $\lvert U_0\rvert = 5$~m/s. The deviations in the phase speeds of the various waves from the zero mean case are considerably larger as expected. Note that the eigenfrequencies of $n=1$ and $n=2$ Rossby modes for the $\beta$-plane system in EE are shown for a smaller range of wavenumbers ($k~{\sim}1$-$15$) owing to the difficulty of resolving the modes at such high shear employing the current numerical method.

\begin{figure}
  \includegraphics[width=1\textwidth]{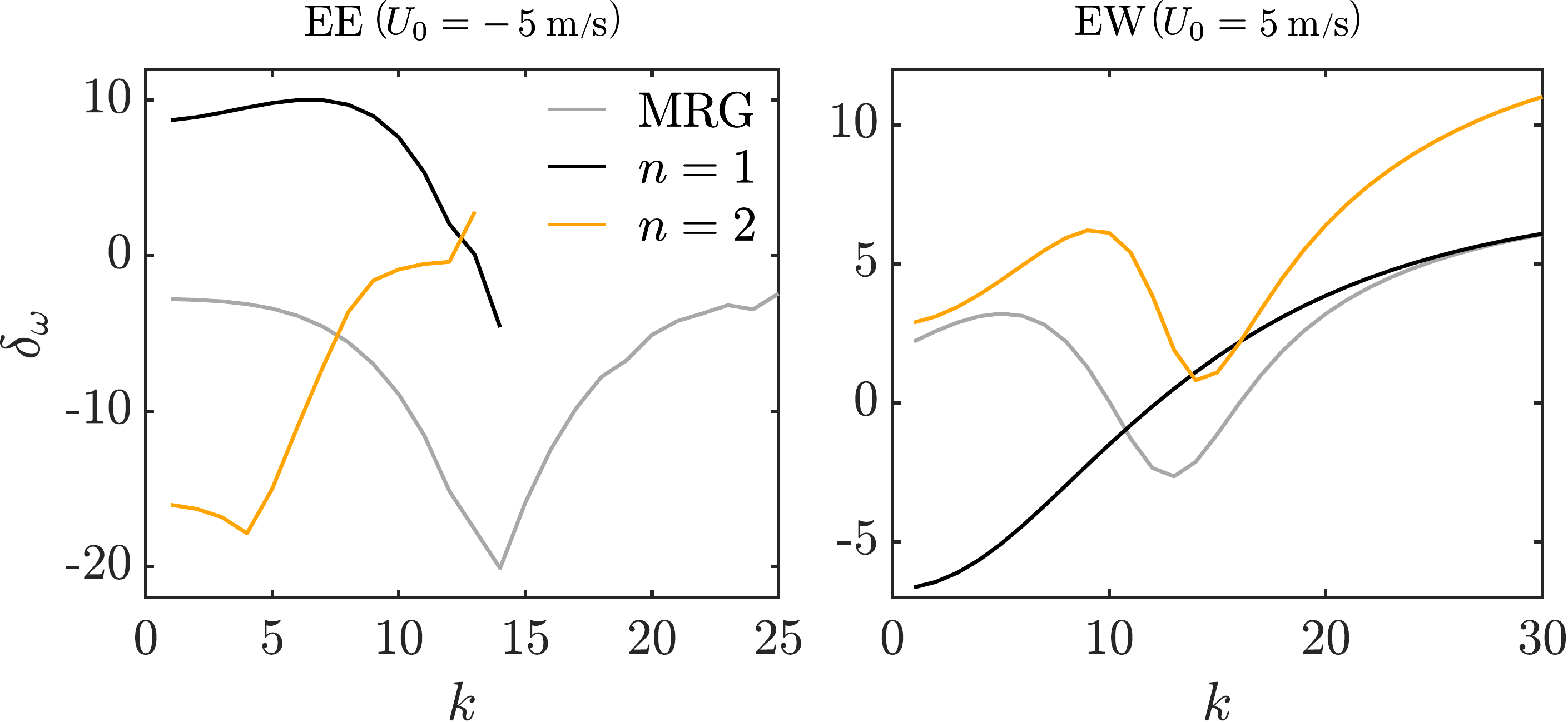}
  \caption{Relative difference $\delta_\omega$ (defined in equation~\eqref{eq:reldiff}) between spectra of the $\beta$-plane and the spherical systems for MRG ($n = 0$) and Rossby modes ($n \ge 1$) for EE (left) and EW (right) when $\lvert U_0\rvert=5$~m/s. The deviations between the two systems are in a similar range as in the low $Fr$ case (see figure \ref{fig:dwr_k}). \label{fig:dwr_k_high_Fr}}
\end{figure}

Beyond these quantitative changes, the system retains most of the qualitative features as observed in the low $Fr$ regime. The non-dispersive nature of short waves in EE is evidently clear even for relatively moderate wavenumbers ($k~{\gtrsim}15$). In EW, as in weak shear, the frequencies of modes corresponding to successive $n$ become nearly indistinguishable with increasing $k$. Compared to the low $Fr$ regime, the clumping of frequencies starts at considerably lower wavenumbers ($k~{\gtrsim}15$). In figure \ref{fig:dwr_k_high_Fr}, we plot $\delta_{\omega}$ as a function of the wavenumber $k$. For the most part, it is seen that the phase speed obtained in the $\beta$-plane is still within $\pm 10\%$ of that of the spherical system. We report that the discrepancy in the phase speeds become more pronounced with increase in $\lvert U_0\rvert$.

\begin{figure}
  \includegraphics[width=1\textwidth]{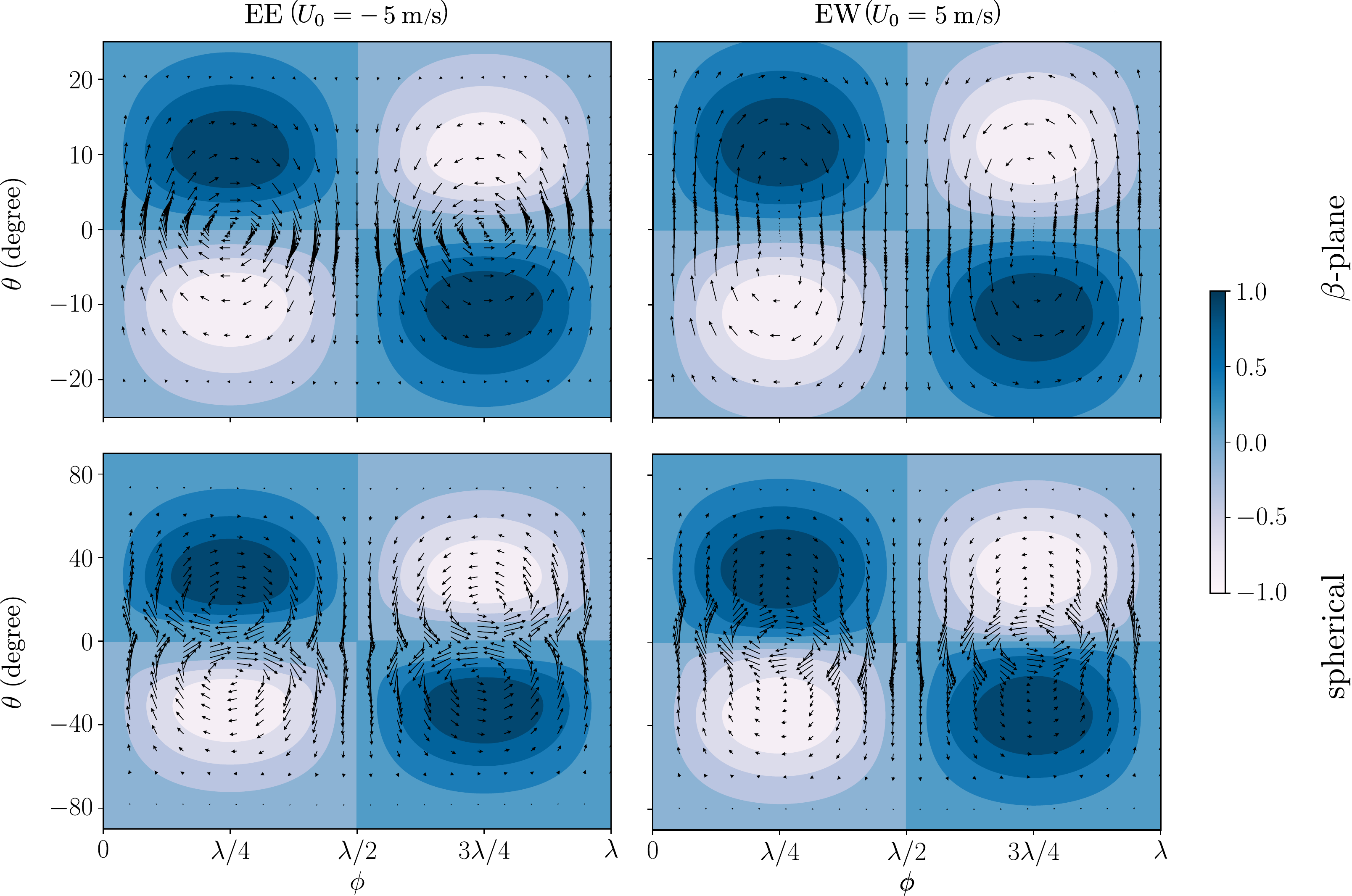}
  \caption{The MRG ($n=0$) modes for $\lvert U_0\rvert = 5~\textrm{m/s}$ and $k=10$ in the $\beta$-plane (top) and spherical (bottom) systems with EE (left) and EW (right); the perturbation wavelength is $\lambda = 2\pi/k = \pi/5$. The arrows depict the velocity and the colour contours depict the surface elevation. Note that the limits in the latitude are different for the top and bottom panels. Compared to low $Fr$ scenarios, there is significantly greater disparity in the degree of trapping of the modes in the two settings.}
  \label{fig:EF_high_Fr_n_0_k_10}
\end{figure}
We next examine the eigenfunction structures in the two settings. These are found to be even more conspicuously different. Figure \ref{fig:EF_high_Fr_n_0_k_10} shows the eigenfunction of MRG mode in the $\beta$-plane and spherical systems for $\lvert U_0\rvert= 5~ \textrm{m/s}$. The eigenfunctions span relatively larger latitudes in both settings compared to the weak shear case. The degree of localisation, however, is significantly different in the two systems. While the eigenfunctions extend to almost the midlatitude in the spherical system (${\sim}5$ times the extent of the mean flow), they extend to only $\lvert\theta\rvert\lesssim 20^{\circ}$ in the $\beta$-plane. On the other hand, the differences in the degree of trapping of the eigenfunctions between EE and EW, which were observed for low $Fr$ case, are no longer as prominent.

\begin{figure}
  \includegraphics[width=\textwidth]{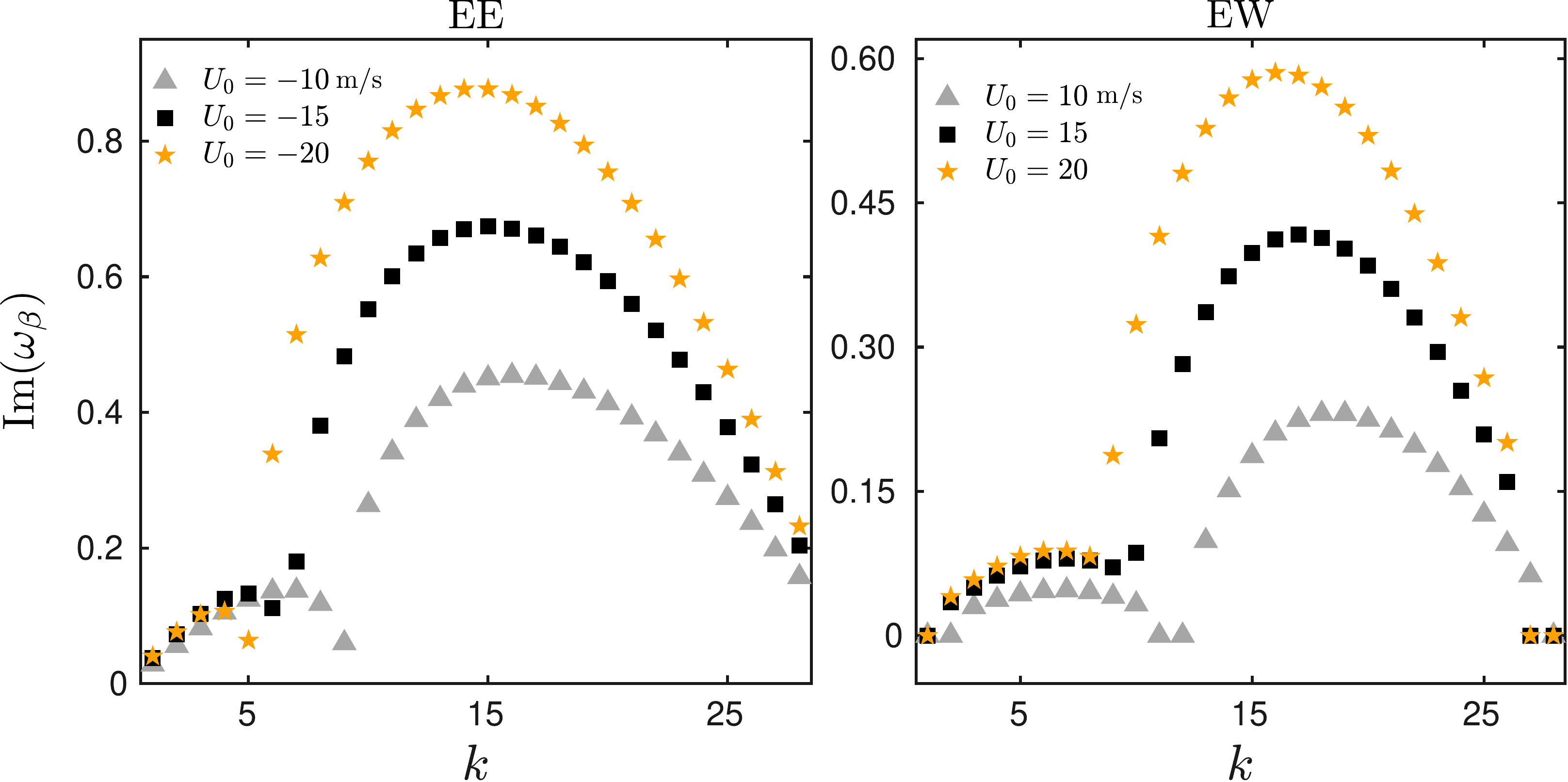}
  \caption{Growth rates of the most unstable mode in the $\beta$-plane setting. The system exhibits higher growth in EE (left) compared to EW (right). The peaks at low and high $k$ correspond to two distinct modes. \label{fig:wi_k_high_Fr}}
\end{figure}

Despite an increase in $Fr$, the base flow remains modally stable in the spherical setting. In stark contrast, for the same mean flow, the $\beta$-plane approximation predicts exponentially growing disturbance modes. Figure~\ref{fig:wi_k_high_Fr} shows the growth rates of the most unstable mode in the $\beta$-plane as a function of the wavenumber $k$ for different values of $\lvert U_0\rvert$; we do not show results for $\lvert U_0\rvert = 5$ m/s as the unstable modes are excited for only a small range of $k$. The non-monotonic nature of the curves arises from the fact that there are different types of unstable modes. For the parameters considered here, at a given $k$, these modes could belong to different branches of the eigenspectrum. As the strength of the mean flow increases, the wavenumber at which the maximum growth rate is achieved reduces.

\begin{figure}
  \includegraphics[width=\textwidth]{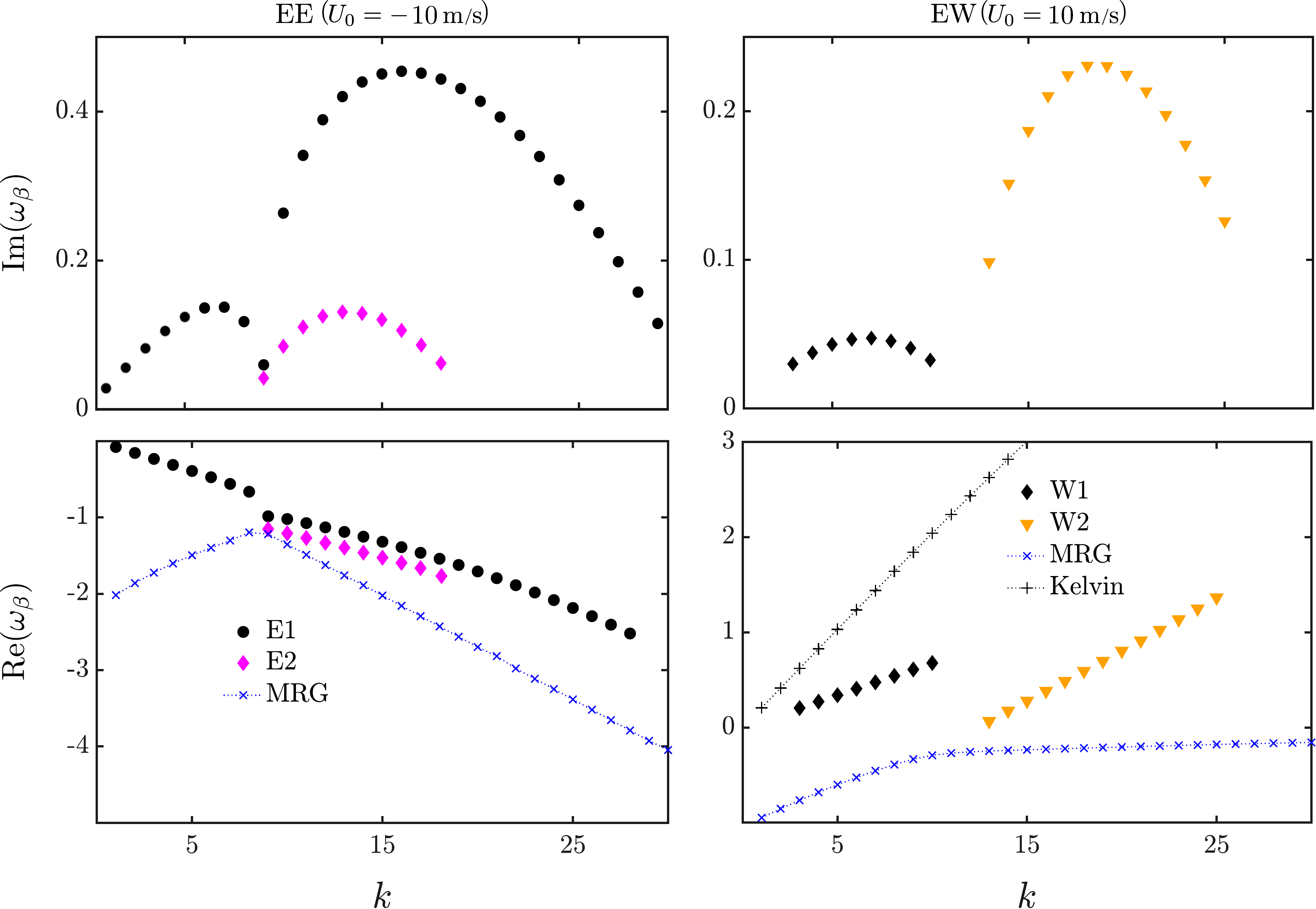}
  \caption{The growth rates (top) and the frequencies (bottom) of different unstable modes of RSWE on $\beta$-plane in the presence of EE (left) and EW (right) for $|U_0|= 10$~m/s. The frequencies of the MRG and Kelvin modes are shown for reference. Note that the spherical system does not yield any unstable modes. \label{fig:w_k_U0_10}}
\end{figure}

Selecting the case with $\lvert U_0\rvert = 10$~m/s, we now take a closer look at the unstable modes obtained in the $\beta$-plane setting. For other $\lvert U_0\rvert$ in this regime, similar behaviour is observed. In figure~\ref{fig:w_k_U0_10}, we show the growth rates $\textrm{Im}(\omega_\beta)$ of the unstable modes and their corresponding frequencies $\textrm{Re}(\omega_\beta)$. One notable feature is that instabilities occur for the lower range of $k$, where the suitability of the $\beta$-plane approximation is already in question. In the bottom panels, we also show the frequencies of the Kelvin and MRG modes. Thus, it is possible to recognise if these modes can be categorised into any of the established wave families.

\begin{figure}
  \includegraphics[width=\textwidth]{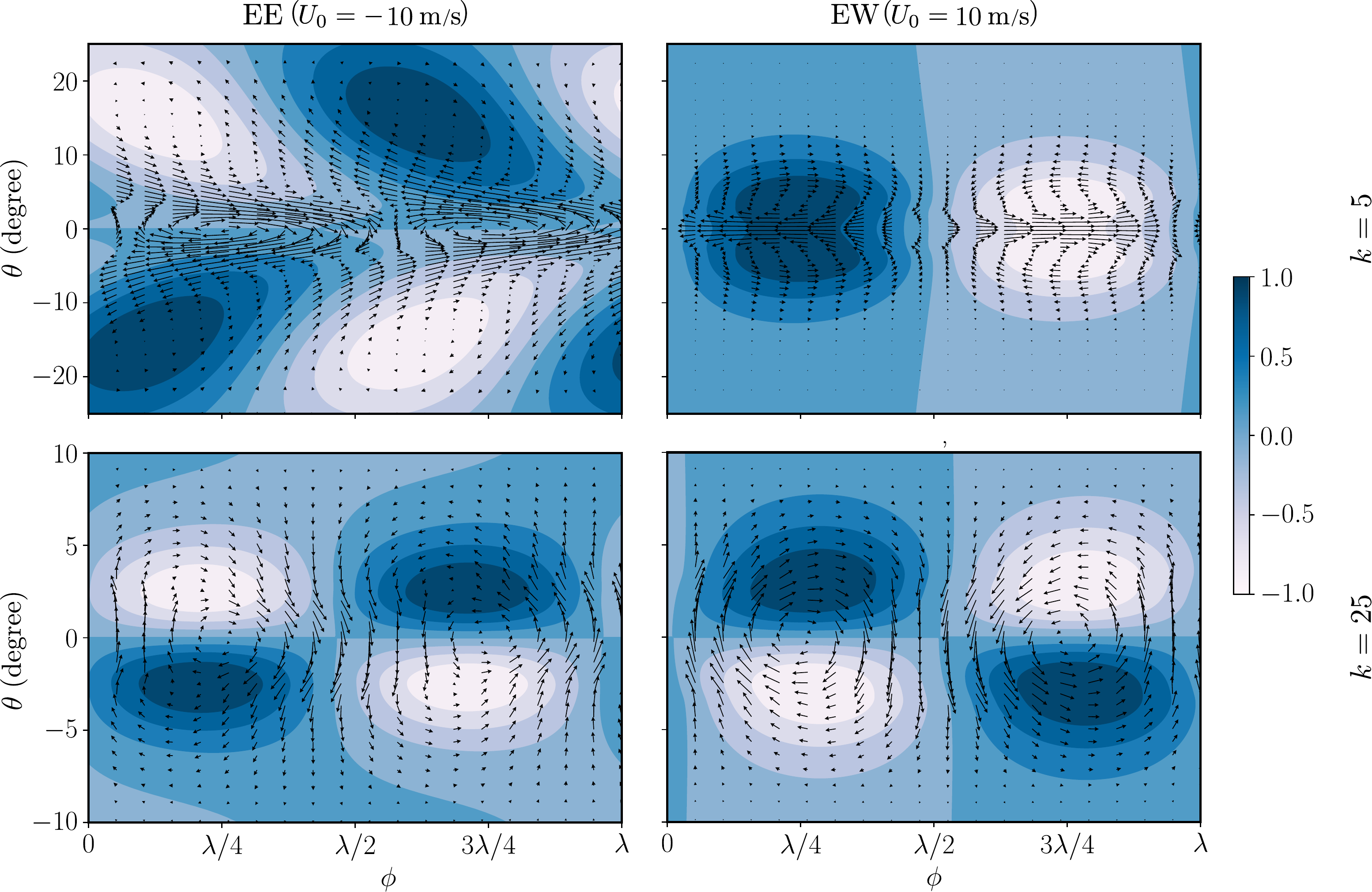}
  \caption{The most unstable mode in the $\beta$-plane setting for $k=5$ (top) and $k=25$ (bottom) in the presence of EE (left) and EW (right) for $\lvert U_0\rvert= 15$~m/s. The arrows depict the velocity and the colour contours depict the surface elevation. Note that the limits in the latitude are different for the top and bottom panels.  \label{fig:EF_high_Fr_most_unst_U0_10}}
\end{figure}

We first consider EE as the mean flow. In this case, the unstable modes are westward-propagating with smaller phase speeds than that of the MRG mode. This suggests these modes have propagation characteristics similar to those of Rossby modes. The corresponding eigenfunctions of the most unstable modes (E1) change gradually with $k$. For low values of $k$, there is an dominant equatorial vortex accompanied by weak off-equatorial vortices (see top left panel in figure \ref{fig:EF_high_Fr_most_unst_U0_10}). As $k$ assumes more moderate values, these two vortical features have comparable strengths. Upon further increase in $k$, a predominantly equatorial vortex resembling the traditional MRG mode emerges (bottom left panel in figure \ref{fig:EF_high_Fr_most_unst_U0_10}). All these modes have even $n$, and consequently the associated geopotential fields are antisymmetric about the equator.
It was shown in figure \ref{fig:wr_k_high_Fr} that there is a range of moderate $k$ for which two unstable modes are found simultaneously. The less dominant unstable mode E2 (not shown) has characterstics of $n = 1$ Rossby mode, i.e., zonal equatorial flows sandwiched between off-equatorial vortices (see the middle panels of figure~\ref{fig:EF_n_1_k_50}). The unstable eigenfunctions are found to be more equatorially trapped with increasing $k$.

On the other hand, with EW as the base flow, the unstable modes are seen to have positive phase velocities. However, these velocities are less than that of the Kelvin wave. Therefore these unstable modes can not be classified into any family identified hitherto. The number of zeros in the meridional velocity component for W1 and W2 are 0 and 1 respectively. Despite these modes propagating eastward, their velocity fields resemble those of the $n = 1$ Rossby at lower $k$ while $n = 0$ MRG mode at large $k$, as shown in the right panel of figure \ref{fig:EF_high_Fr_most_unst_U0_10}. As in the case with EE as the base flow, the unstable eigenfunctions become more localised around the equator as $k$ increases.

It is apparent that these unstable modes in $\beta$-plane discussed above have no counterpart in the spherical system. They emerge as an artefact of the $\beta$-plane approximation, and may be termed spurious for the earth system. This calls for caution while employing the $\beta$-plane approximation for stability studies, even for highly equatorially confined mean flows.

\section{Summary and discussion}\label{sec:Summary}
In this study, the stability characteristics of confined equatorial mean shear flows have been examined in the $\beta$-plane and spherical coordinate systems. It is seen that shear flow, even while itself highly confined to small latitudes, causes the perturbation energy to spread meridionally to a major extent, at all wavenumbers, and for all families of atmospheric waves. It is hoped that this finding will motivate detailed simulations and comparisons with observations to understand non-local effects of shear flow on the earth system.

The $\beta$-plane approximation is designed to simplify the study of equatorial dynamics and is expected to correlate well with that of the full spherical system when the flow is confined near the equator. However we find that even under these conditions which are seemingly favourable for comparison, there are several notable differences in the results obtained with the linear $\beta$-plane and spherical systems, both in the spectra and the spatial structure of eigenmodes. Hence, in the presence of any equatorial shear, the stability analysis of the atmosphere must be performed in the full spherical system.

When the mean flow is weak with low $Fr$ (section~\ref{sec:LowFr}), the modes that are prominently affected are the MRG and Rossby modes. These waves propagate faster (slower) in the presence of EE (EW) when compared with the zero mean flow case. With respect to the zero mean flow configuration, the eigenfunctions are more (less) equatorially trapped in EE (EW). The eigenfunctions are highly equatorially confined in the $\beta$-plane setting. With EE as the mean flow, at high values of the wavenumber $k$, the waves become non-dispersive with a phase speed lower than that of the Kelvin mode. With EW, at high wavenumbers, the phase speeds of modes with successive $n$ become increasingly close to each other. The consequences of this behaviour are reflected in the topological properties of eigenfunctions. The equatorially centred vortex (a signature of modes with even $n$ in the absence of shear) bifurcates into two off-equatorial vortices for higher $k$. This finding would be important to explore in the context of observations.

For stronger mean flows with larger $Fr$ (section~\ref{sec:High_Fr}), the dispersion curves in the two settings are qualitatively different, with additional branches appearing in the spectrum under the $\beta$-plane approximation. The physical origin of these branches and their relevance to equatorial dynamics would be important avenues for future research. More importantly, while the flow is shown to be unstable, with exponentially growing modes, in the $\beta$-plane system, the spherical system continues to remain neutrally stable. We note that these instabilities are seen only for a lower range of wavenumbers that are ill-suited for analysis in the $\beta$-plane, but which nevertheless are sometimes studied under this approximation. On analysing the stability of the mean flow as a function of $k$, it is found that synoptic scale waves ($\lambda \sim 2500$~km) exhibit largest growth rates in most of the cases.

The methods used and the assumptions made in the present study naturally present some limitations to their application in a more realistic setting. The shallow water approximation neglects the variation of the vertical structure of atmosphere. In addition to this, the effects of a longitudinally varying mean flow and a non-zero meridional mean flow have not been studied here; such a scenario could also involve significant variations of the equivalent depth of the shallow layer. With effects of stratification (or vertical structure) incorporated, \cite{killworth1980} showed that there are different types of instabilities depending on the lateral extent of the mean flow. It would be interesting to also examine if different classes of unstable modes emerge in a single layer as the mean flow's extent is varied. Finally, the analysis is linear, which restricts the investigation into the system over time scales when the perturbations are small enough for linearity to hold. However, the results obtained even with such a simplified model have non-trivial implications for the theoretical understanding of equatorial waves.

As discussed in~\cite{WZ89}, the smaller degree of equatorial trapping of the Rossby waves in EW can be thought of as leading to an enhanced interaction between the tropics and extratropics. Our study shows that all the modes are significantly less trapped about the equator in the spherical system. The $\beta$-plane system thus severely underestimates the degree of tropical-extratropical interaction facilitated by Rossby waves. Further, the defining features of the Rossby waves corresponding to even values of $n$, i.e., cross equatorial flow with an equatorially centred vortex cease to hold in EW at large $k$. The eigenfunctions with off-equatorial vortices and a very weak equatorial flow can thus serve as markers for the identification of mesoscale ($\lambda\sim 500$~km) Rossby waves in the atmospheric in the presence of a westerly mean flow.

\appendix
\section{Effects of varying equivalent depth \texorpdfstring{$H_0$}{H0}}
\label{ap:var_H0}
The effects of varying the equivalent depth $H_0$ are presented here for the Gaussian mean flows considered with $\lvert U_0 \rvert = 0.5$ m/s. Note that varying $H_0$ while keeping $U_0$ fixed is effectively changing the value of $Fr$, i.e., increasing $H_0$ results in lower value of $Fr$ and vice versa.

\begin{figure}
  \includegraphics[width=\textwidth]{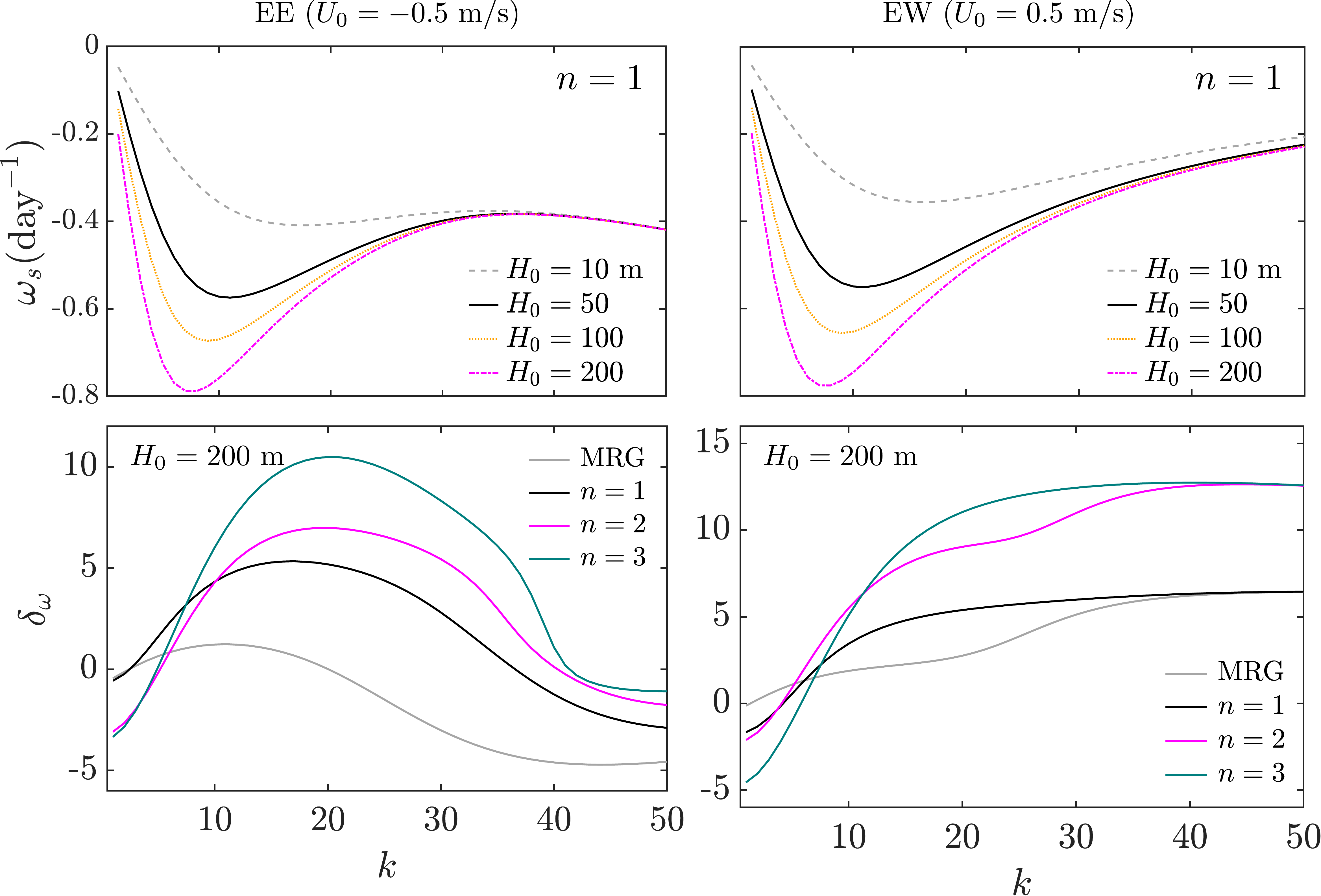}
  \caption{Top panels: dispersion curves for $n=1$ Rossby mode in the spherical system in the presence of $EE$ (left) and $EE$ (right) for different values of equivalent depth ($H_0$).
Bottom panels: Relative difference ($\delta_\omega$) (defined in eq. \ref{eq:reldiff}) between the spectra of the $\beta$-plane and the spherical system for $MRG$ (n=0) and Rossby modes ($n>1$) for $H_0=200$ m in the presence of EE (left) and EW (right). \label{fig:wr_k_low_Fr_diff_H_0}}
\end{figure}
  
The top panels of figure \ref{fig:wr_k_low_Fr_diff_H_0} show the dispersion curves for a $n=1$ Rosby mode in EE (left) and EW (right) for different values of $H_0$ in the spherical system. For a range of moderate wavenumbers ($k\sim 1-30$), the frequency increases with increasing $H_0$. 
Curiously, the frequencies of short waves ($k\gtrsim 40$), appear to be only weakly sensitive to the value of $H_0$.
We have verified that these features hold for other Rossby modes with different $n$. Similar dispersion curves are also obtained when the analysis is performed using the $\beta$-plane approximation. The bottom panels of the figure \ref{fig:wr_k_low_Fr_diff_H_0} shows $\delta_{\omega}$ as a function of $k$ when $H_0=200$ m. While the deviations remain within nearly $10\%$ when EE is the mean flow, $\delta_{\omega}$ takes on relatively larger values for EW. This is similar to what we have earlier for $H_0=100$ m (see figure \ref{fig:dwr_k}).

\begin{figure}
  \includegraphics[width=\textwidth]{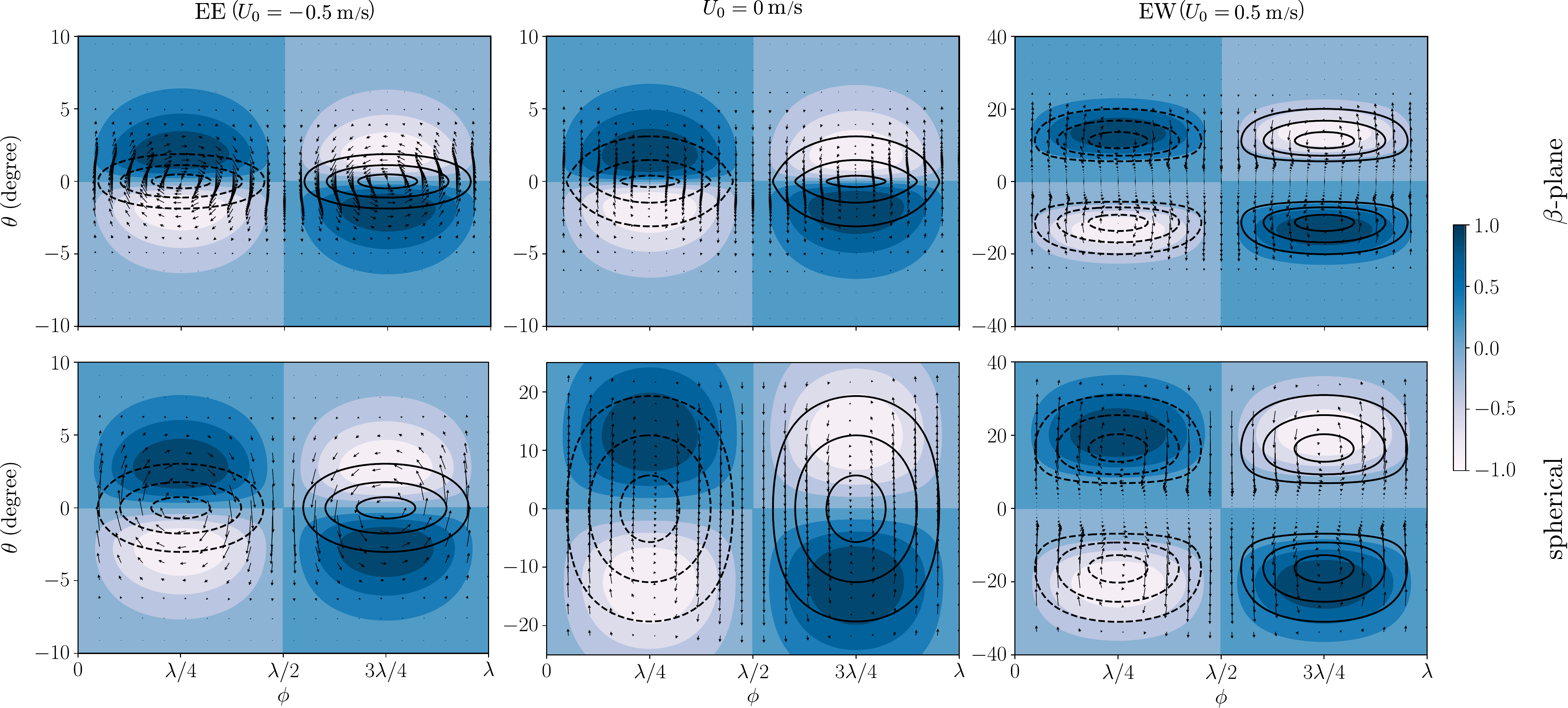}
  \caption{Same as figure \ref{fig:EF_n_0_k_50}, but for $H_0 = 200$ m. The eigenfunctions are observed to be more trapped for a higher $H_0$.}
  \label{fig:EF_n_0_k_50 H0_200}
\end{figure}

We plot the MRG mode for $H_0 = 200$ m and $\lvert U_0\rvert = 0.5$ m/s in figure \ref{fig:EF_n_0_k_50 H0_200}. It is seen that the eigenfunctions tend to be more equatorially trapped in both the $\beta$-plane and spherical settings. As was seen earlier, the eigenfunction in the spherical system is seen to have a greater extent beyond the equator than its counterpart in the $\beta$-plane.

\section{Stability analysis of Mexican hat profiles}
\label{ap:mex_hat}
\begin{figure}
\centering
\includegraphics[width=0.55\textwidth]{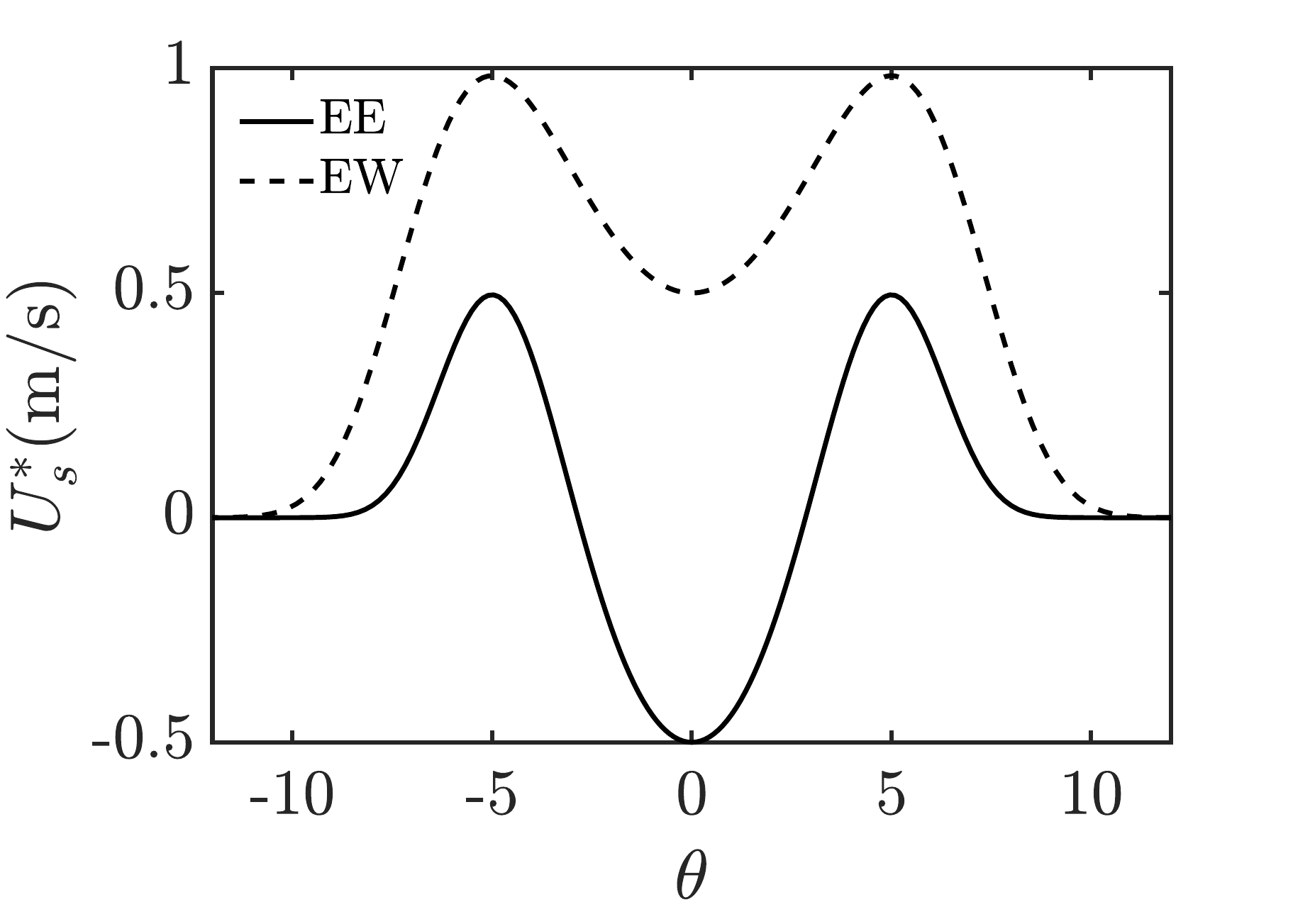}
\caption{Profiles of equatorial easterly (EE) and equatorial westerly (EW) mean flow with Mexican hat profiles.\label{fig:BF mexhat}}
\end{figure}

Here, we analyse a different mean flow -- one with a Mexican hat profile (see figure \ref{fig:BF mexhat}). This flow is characterised off-equatorial westerly flow; at the equator, the mean flow velocity could be negative (corresponding to EE) or positive (corresponding to EW). These profiles are motivated by those considered by \citet{WZ89}. 
  Compared to their case, we choose the profiles to span considerably smaller latitudes so as to provide us with conditions which are favourable for a comparison between the results of the $\beta$-plane and spherical system. Further, the mean shear considered here is much weaker. For the velocity at the equator, we choose $\lvert U_0 \rvert = 0.5$ m/s. As before, the mean flow is labelled as EE or EW depending on the sign of $\lvert U_0 \rvert$.

\begin{figure}
  \includegraphics[width=\textwidth]{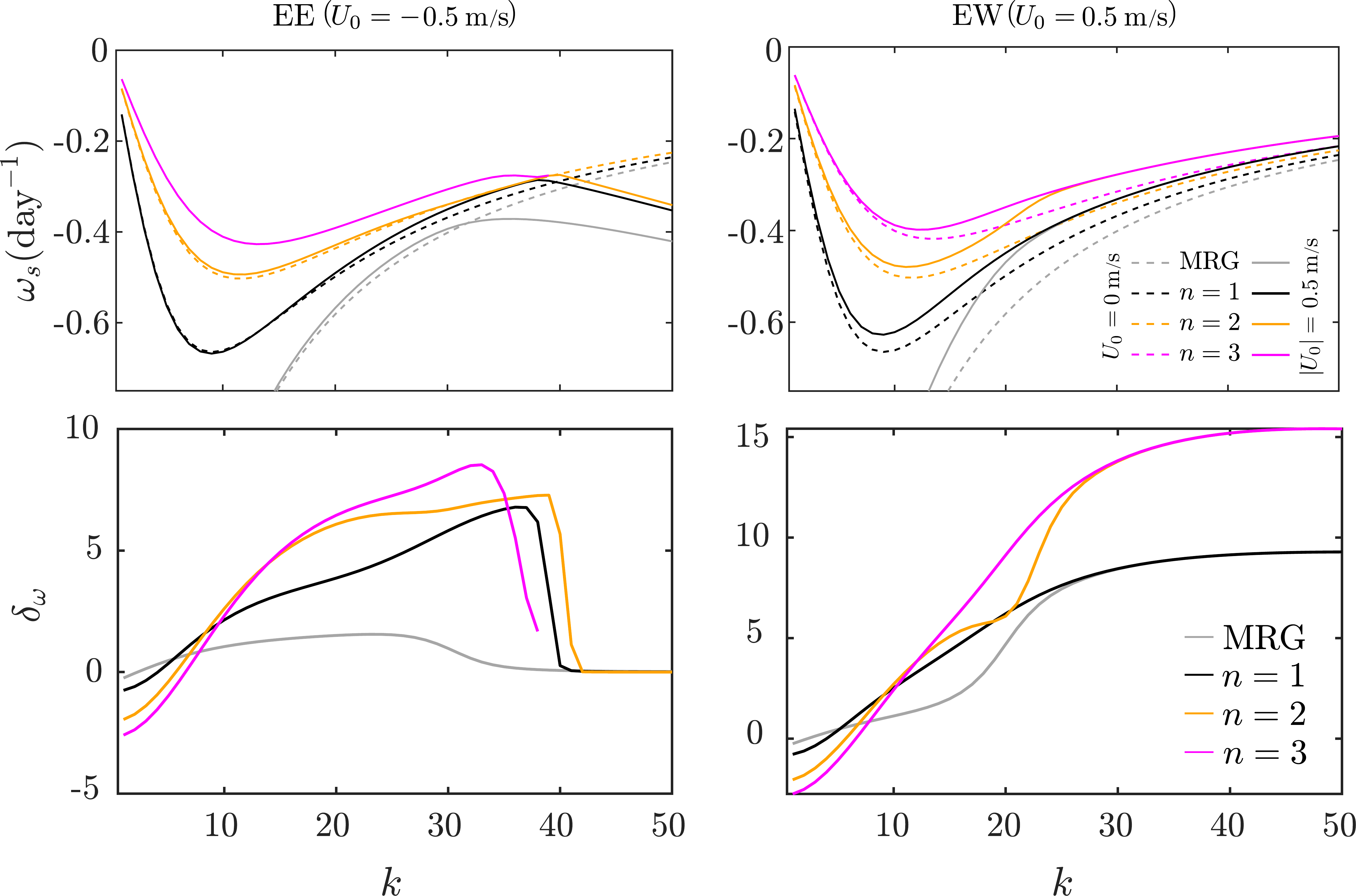}
  \caption{Top panels: dispersion curves for the first few Rossby and MRG waves in the spherical system (top) for EE (left) and EW (right) with $\lvert U_0\rvert=0.5$~m/s ($\lvert Fr\rvert=0.02$); the no mean flow case is shown for a reference. Bottom panels: $\delta_\omega$ for the MRG ($n = 0$) and Rossby modes ($n \ge 1$) for EE (left) and EW (right).
   \label{fig:wr_k_low_Fr mexhat}}
\end{figure}

The top panels of figure \ref{fig:wr_k_low_Fr mexhat} show the dispersion curves for MRG ($n = 0$) and Rossby modes for the zero shear case and when $\lvert U_0\rvert = 0.5$ m/s. The qualitative features observed when the mean flow took the form of the Gaussian profile continue to hold. Short waves (large $k$) become nearly non dispersive when the base flow is EE. With EW as the base flow, the frequencies of short waves for modes with successive $n$ become nearly indistinguishable. The bottom panels show the relative difference in the frequencies of these modes in the $\beta$-plane and spherical system. For EE, $\delta_{\omega}$ is seen to be within 10 \%. On the other hand, with EW as the mean flow, $\delta_{\omega}$ takes values that are even higher than those for EW with a Gaussian profile. As in Gaussian profiles (see figure \ref{fig:dwr_k}), the differences in frequency is relatively larger in EW than in EE.

\begin{figure}
  \includegraphics[width=\textwidth]{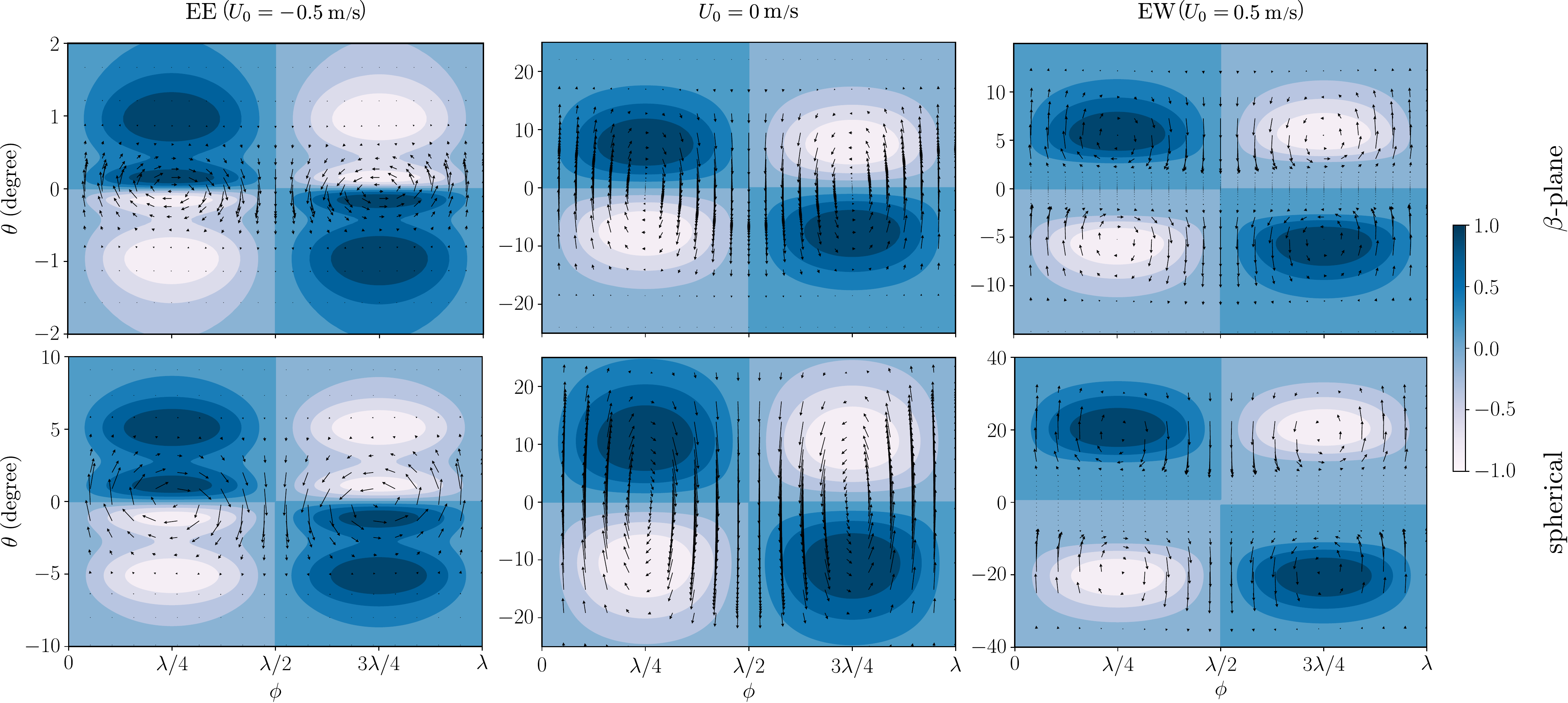}
  \caption{Same as figure \ref{fig:EF_n_0_k_50}, but for Mexican hat profile with $\lvert U_0\rvert = 0.5$ m/s. For this mode, $k = 50$ and $n = 0$. The broader qualitative features observed for the Gaussian profile continue to hold even for the Mexican hat mean flow.}
  \label{fig:EF_n_0_k_50 mexhat}
\end{figure}

Figure \ref{fig:EF_n_0_k_50 mexhat} shows the eigenfunctions of the MRG mode ($n = 0$) obtained in the $\beta$-plane (left) and spherical (right) settings for EW when $k = 50$. When comparing with the eigenfunction shown in figure \ref{fig:EF_n_0_k_50}, we see a greater degree of localisation of the eigenfunction when the mean flow assumes the Mexican hat profile. As in the case with the Gaussian mean flow profile, the eigenfunctions are considerably more trapped in the $\beta$-plane system. We have verified that the qualitative features seen for the Gaussian mean flow is observed for the Mexican hat mean flow profile for other Rossby modes as well.

\section{Perturbation zonal velocity and geopotential}
\label{ap:uh_v}
Using the eigenvalue problems defined by equations \ref{Eq:EVP_s} and \ref{Eq:EVP_b} as a starting point, it is possible to obtain expressions for any component of the eigenvector in terms of another component. In this work, the meridional velocity of the mode is employed to identify its family. Here we provide expressions for the zonal perturbation velocity and the perturbation height in terms of the meridional perturbation velocity. 

\subsection{Spherical coordinate system}
From equation \ref{Eq:EVP_s}, we get:
\begin{align}
  \tilde{u}_{s} = i\mathcal{S}_{uv}\tilde{v}_{s} + i\mathcal{S}_{uDv}D_y\tilde{v}_{s},~\tilde{h}_{s} = i\mathcal{S}_{hv}\tilde{v}_{s} + i\mathcal{S}_{hDv}D_y\tilde{v}_{s}
\end{align}
where
\begin{subequations}\label{Eq:uh_coeff_s}
  \begin{align}
  &\mathcal{S}_{uv} = \frac{\epsilon\hat{\omega}_s\Omega_{as}\cos\theta - kD_{\theta    }\left(\mathcal{H}_s\cos\theta\right)}{\epsilon\hat{\omega}_s^2 - k^2\mathcal{H}_s}~, \\
  &\mathcal{S}_{uDv} = -\frac{k\mathcal{H}_s\cos\theta}{\epsilon\hat{\omega}_s^2 - k^2\mathcal{H}_s}~, \\
  &\mathcal{S}_{hv} = \frac{\epsilon\left[k\mathcal{H}_s\Omega_{as}\cos\theta - \hat{\omega}_sD_{\theta}\left(\mathcal{H}_s\cos\theta\right)\right]}{\epsilon\hat{\omega}_s^2 - k^2\mathcal{H}_s}~, \\
  &\mathcal{S}_{hDv} = -\frac{\epsilon\hat{\omega}_s\mathcal{H}_s\cos\theta}{\epsilon\hat{\omega}_s^2 - k^2\mathcal{H}_s}~.
\end{align}
\end{subequations}
In the above expressions, $\hat{\omega}_s = \omega_s\cos\theta - kU_s$.

\subsection{\texorpdfstring{$\beta$}{beta}-plane system}
From equation \ref{Eq:EVP_b}, we get:
\begin{align}
  \tilde{u}_{\beta} = i\mathcal{B}_{uv}\tilde{v}_{\beta} + i\mathcal{B}_{uDv}D_y\tilde{v}_{\beta},~\tilde{h}_{\beta} = i\mathcal{B}_{hv}\tilde{v}_{\beta} + i\mathcal{B}_{hDv}D_y\tilde{v}_{\beta},
\end{align}
where
\begin{subequations}\label{Eq:uh_coeff_b}
  \begin{align}
    &\mathcal{B}_{uv} = \frac{k_{\beta}yU_{\beta} + \left(\omega_{\beta} - k_{\beta}U_{\beta}\right)\left(y - D_yU_{\beta}\right)}{\left(\omega_{\beta} - k_{\beta}U_{\beta}\right)^2 - k_{\beta}^2\mathcal{H}_\beta}~, \\
    &\mathcal{B}_{uDv} = -\frac{k_{\beta}\mathcal{H}_\beta}{\left(\omega_{\beta} - k_{\beta}U_{\beta}\right)^2 - k_{\beta}^2\mathcal{H}_\beta}~, \\
    &\mathcal{B}_{hv} = \frac{yU_{\beta}\left(\omega_{\beta} - k_{\beta}U_{\beta}\right) + k_{\beta}\mathcal{H}_\beta\left(y - D_yU_{\beta}\right)}{\left(\omega_{\beta} - k_{\beta}U_{\beta}\right)^2 - k_{\beta}^2\mathcal{H}_\beta}~, \\
    &\mathcal{B}_{hDv} = -\frac{\mathcal{H}_\beta\left(\omega_{\beta} - k_{\beta}U_{\beta}\right)}{\left(\omega_{\beta} - k_{\beta}U_{\beta}\right)^2 - k_{\beta}^2\mathcal{H}_\beta}~.
  \end{align}
\end{subequations}
When $U_{\beta} = 0$ (and hence $\mathcal{H}_{\beta} = 1$), the expressions for $\tilde{u}_{\beta}$ and $\tilde{h}_{\beta}$ reduce to those given in \cite[section 8.2]{vallis_2017}.

\bibliographystyle{elsarticle-harv} 
\bibliography{ref_stab_analys}

\end{document}